\documentclass[longauth]{aa}

 \usepackage{natbib}
\bibpunct{(}{)}{;}{a}{}{,} 
\usepackage{graphicx}
\usepackage{txfonts}
\usepackage{multirow}
\usepackage{adjustbox}
\usepackage{xcolor}
\definecolor{cadmiumgreen}{rgb}{0.0, 0.7, 0.2}
\usepackage{hyperref}
\hypersetup{
  colorlinks = true,
  citecolor = cadmiumgreen
}
\begin{document}

\title{A novel framework for semi-Bayesian radial velocities through template matching}

\author{A. M. Silva\inst{\ref{Inst_1}, \ref{Inst_2}}
        \and J. P. Faria\inst{\ref{Inst_1},\ref{Inst_2}} 
        \and N. C. Santos\inst{\ref{Inst_1},\ref{Inst_2}}
        \and S. G. Sousa\inst{\ref{Inst_1}} 
        \and P. T. P. Viana \inst{\ref{Inst_1},\ref{Inst_2}}
        \and J. H. C. Martins\inst{\ref{Inst_1}} 
        \and P. Figueira\inst{\ref{Inst_7},\ref{Inst_1}} 
        \and C. Lovis\inst{\ref{Inst_5}} 
        \and F. Pepe\inst{\ref{Inst_5}} 
        \and S. Cristiani\inst{\ref{Inst_11}} 
        \and R. Rebolo \inst{\ref{Inst_3},\ref{Inst_4}, \ref{Inst_12}}
        \and R. Allart\inst{\ref{Inst_20}, \ref{Inst_5}} 
        \and A. Cabral\inst{\ref{Inst_15},\ref{Inst_17}} 
        \and A. Mehner\inst{\ref{Inst_7}} 
        \and A. Sozzetti\inst{\ref{Inst_9}} 
        \and  A. Suárez Mascare\~no\inst{\ref{Inst_3},\ref{Inst_4}} 
        \and C. J.A.P. Martins\inst{\ref{Inst_1}, \ref{Inst_19}} %
        \and D. Ehrenreich\inst{\ref{Inst_5}} 
        \and D. M{\'e}gevand\inst{\ref{Inst_5}} 
        \and E. Palle\inst{\ref{Inst_3}, \ref{Inst_4}}
        \and G. Lo Curto\inst{\ref{Inst_7}} 
        \and H. M. Tabernero\inst{\ref{Inst_10}}  %
        \and J. Lillo-Box\inst{\ref{Inst_21}} 
        \and J. I. Gonz\'alez Hern\'andez\inst{\ref{Inst_3}, \ref{Inst_4}} 
        \and M. R.  Zapatero Osorio\inst{\ref{Inst_10}} 
        \and N. C. Hara\inst{\ref{Inst_5}}  
        \and N. J. Nunes\inst{\ref{Inst_15}} 
        \and P. Di Marcantonio\inst{\ref{Inst_11}} %
        \and S. Udry\inst{\ref{Inst_5}}
        \and V. Adibekyan\inst{\ref{Inst_1},\ref{Inst_2}} 
        \and X. Dumusque\inst{\ref{Inst_5}} 
         }
\institute{
          Instituto de Astrof\'isica e Ci\^encias do Espa\c{c}o, Universidade do Porto, CAUP, Rua das Estrelas, 4150-762 Porto, Portugal \label{Inst_1} 
          \and
          Departamento de F\'isica e Astronomia, Faculdade de Ci\^encias, Universidade do Porto, Rua do Campo Alegre, 4169-007 Porto, Portugal \label{Inst_2}
          \and
          Instituto de Astrof\'{i}sica de Canarias (IAC), 38205 La Laguna, Tenerife, Spain \label{Inst_3}
          \and
          Universidad de La Laguna (ULL), Departamento de Astrof\'{i}sica, 38206 La Laguna, Tenerife, Spain \label{Inst_4}
         \and
         Département d’astronomie de l’Universit\'e de Gen\`eve, Chemin Pegasi 51, 1290 Versoix, Switzerland \label{Inst_5}
         \and
         INAF - Osservatorio Astronomico di Brera, Via Bianchi 46, 23807 Merate, Italy \label{Inst_6} 
         \and
         European Southern Observatory, Alonso de C\'ordova 3107, Vitacura, Regi\'on Metropolitana, Chile\label{Inst_7} 
        \and
        INAF - Osservatorio Astronomico di Palermo, Piazza del Parlamento 1, 90134 Palermo, Italy\label{Inst_8}
        \and
        INAF - Osservatorio Astrofisico di Torino, via Osservatorio 20, 10025 Pino Torinese, Italy \label{Inst_9}
        \and
        Centro de Astrobiolog\'\i a (CSIC-INTA), Crta. Ajalvir km 4, E-28850 Torrej\'on de Ardoz, Madrid, Spain \label{Inst_10} 
        \and
        INAF - Osservatorio Astronomico di Trieste, via G. B. Tiepolo 11, I-34143 Trieste, Italy \label{Inst_11} 
        \and
        Consejo Superior de Investigaciones Cient\'{\i}cas, Spain\label{Inst_12}
        \and
        Physics Institute, University of Bern, Sidlerstrasse 5, 3012 Bern, Switzerland\label{Inst_13}
        \and
        Institute for Fundamental Physics of the Universe, Via Beirut 2, I-34151 Grignano, Trieste, Italy\label{Inst_14}
        \and
        Instituto de Astrof\'isica e Ci\^encias do Espa\c{c}o, Faculdade de Ci\^encias da Universidade de Lisboa, Campo Grande, PT1749-016 Lisboa, Portugal \label{Inst_15}
        \and
        Fundaci\'on G. Galilei -- INAF (Telescopio Nazionale Galileo), Rambla J. A. Fern\'andez P\'erez 7, E-38712 Bre\~na Baja, La Palma, Spain\label{Inst_16} 
        \and
        Faculdade de Ci\^encias da Universidade de Lisboa (Departamento de F\'isica), Edif\'icio C8, 1749-016 Lisboa, Portugal \label{Inst_17}
        \and
        European Southern Observatory, Karl-Schwarzschild-Strasse 2, 85748  Garching b. M\"unchen, Germany \label{Inst_18}
        \and
        Centro de Astrof\'isica da Universidade do Porto, Rua das Estrelas, 4150-762 Porto, Portugal\label{Inst_19}
        \and
        Department of Physics, and Institute for Research on Exoplanets, Universit\'e de Montr\'eal, Montr\'eal, H3T 1J4, Canada \label{Inst_20}
        \and
         Centro de Astrobiolog\'ia (CAB, CSIC-INTA), Depto. de Astrof\'isica, ESAC campus, 28692, Villanueva de la Ca\~nada (Madrid), Spain \label{Inst_21}
}

\date{Received date / Accepted date }


\abstract
{The detection and characterization of an increasing variety of exoplanets has been in part possible thanks to the continuous development of high-resolution, stable spectrographs, and using the Doppler radial-velocity (RV) method. The Cross Correlation Function (CCF) method is one of the traditional approaches for the derivation of RVs. More recently, template matching has been introduced as an advantageous alternative for M-dwarf stars.}
{We describe a new implementation of the template matching technique for stellar RV estimation within a semi-Bayesian framework, providing a more statistically principled characterization of the RV measurements and associated uncertainties. This methodology, named S-BART: Semi-Bayesian Approach for RVs with Template-matching, can currently be applied to HARPS and ESPRESSO data. We first validate its performance with respect to other template matching pipelines using HARPS data. Then, we apply \texttt{S-BART} to ESPRESSO observations, comparing the scatter and uncertainty of the derived RV time series with those obtained through the CCF method. We leave, for future work, a full analysis of the planetary and activity signals present in the datasets considered.}
{In the context of a semi-Bayesian framework, a common RV shift is assumed to describe the difference between each spectral order of a given stellar spectrum and a template built from the available observations. Posterior probability distributions are obtained for the relative RV associated with each spectrum using the Laplace approximation, after marginalization with respect to the continuum. We also implemented, for validation purposes, a traditional template matching approach, where a RV shift is estimated individually for each spectral order and the final RV estimate is calculated as a weighted average of the individual order's RVs.}
{The application of our template-based methods to HARPS archival observations of Barnard’s star allowed us to validate our implementation against other template matching methods. Although we found similar results, the RMS of the RVs derived with \texttt{S-BART} was smaller than that obtained with the \texttt{HARPS-TERRA} and \texttt{SERVAL} pipelines. We believe this is due to differences in the construction of the stellar template and the handling of telluric features. After validating \texttt{S-BART}, we applied it to 33 ESPRESSO GTO targets, evaluating its performance and comparing it with respect to the CCF method as implemented in the ESO pipeline. We found a decrease in the median RV scatter of $\sim$10\% and $\sim$4\% for M- and K-type stars, respectively. Our semi-Bayesian framework yields more precise RV estimates than the CCF method, in particular in the case of M-type stars where \texttt{S-BART} achieves a median uncertainty of $\sim$ 15 ${\rm cm}\ {\rm s}^{-1}$ over 309 observations of 16 targets. Further, with the same data we estimated the nightly zero point (NZP) of the instrument, finding a weighted NZP scatter below $\sim$ 0.7 ${\rm m}\ {\rm s}^{-1}$. Given that this includes stellar variability, photon noise, and potential planetary signals, it should be taken as an upper limit of the RV precision attainable with ESPRESSO data.}
{}

\maketitle
%

%

\section{Introduction}

Finding and characterizing other Earths -- rocky planets with the physical conditions to hold liquid water on their surface -- is one of the boldest goals of present-day astrophysics. The discovery \citep[e.g.][]{mayorHARPSSearchSouthern2011, mayorDopplerSpectroscopyPath2014, hsuOccurrenceRatesPlanets2019, rosenthalCaliforniaLegacySurvey2021} that rocky planets are actually very common around solar-type stars, i.e. late F, G and early K stars, made this goal more achievable and motivated the development of a new generation of ground and space-based instruments and missions (e.g. ESPRESSO - \citealt{pepeESPRESSOVLTOnsky2021}, PLATO - \citealt{rauerPLATOMission2014}, HIRES@ELT - \citealt{marconiHIRESHighresolutionSpectrograph2021}).

One of the most prolific exoplanet discovery method is the radial velocity (RV) method, based on the detection of variations in the velocity of a star along our line of sight, induced by the gravitational pull of planetary companions. However, the identification of Earth-like planets, orbiting solar-type stars, poses a significant challenge: Earth itself induces a signal with an amplitude of only \textasciitilde 9 ${\rm cm}\ {\rm s}^{-1}$ on the Sun. In order to achieve this RV precision domain a new generation of spectrographs have been developed. An example of a state-of-the-art spectrograph is ESPRESSO, the “Échelle SPectrograph for Rocky Exoplanets and Stable Spectroscopic Observations”, built to reach a precision of 10 ${\rm cm}\ {\rm s}^{-1}$ with a wavelength coverage from 380 to 788 nm \citep{pepeESPRESSOVLTOnsky2021}.

The first confirmed detection of an exoplanet around a solar-type star, \textit{51 Pegasi b} \citep{mayorJupitermassCompanionSolartype1995}, was achieved with radial velocities computed using the Cross Correlation Function (CCF) method. In the method's early stages, a binary mask, with  fixed non-zero weights attributed to the expected positions of stellar absorption lines, was cross-correlated with the spectra \citep{baranneELODIESpectrographAccurate1996}. However, as deep sharp lines contain more information than broad and shallower ones \citep[as shown by the methodology introduced in][]{bouchyFundamentalPhotonNoise2001} the masks were improved by associating different weights to different lines \citep{pepeCORALIESurveySouthern2002}, depending on the RV information present in them, such that deep lines contribute more to the final CCF profile than shallow ones.

Even though the CCF method has been widely used, building the masks can be a challenging task in some situations, especially for M dwarfs \citep[e.g.][]{rainerStellarMasksBisector2020,lafargaCARMENESSearchExoplanets2020a}. The high number of stellar spectral lines, due to the lower temperatures of M-type stars, results in the presence of a larger number of spectral lines, with most of them being spectroscopically blended, hardening the construction of the CCF mask and the fitting of the CCF profile. For such cases it has been shown that template matching can surpass the CCF method \citep[e.g.][]{anglada_escude_HARPS_TERRA_2012,zechmeisterSpectrumRadialVelocity2018,lafargaCARMENESSearchExoplanets2020a}, as the stellar template will contain a large majority of the lines in the stellar spectrum. This is a data-driven method, where each spectra is compared against a template built from available observations, and has been implemented in \texttt{HARPS-TERRA} \citep{anglada_escude_HARPS_TERRA_2012}, \texttt{NAIRA} \citep{astudillo-defruSearchEarthlikePlanets2015} and \texttt{SERVAL} \citep{zechmeisterSpectrumRadialVelocity2018}. More recently, new approaches to RV estimation have emerged, based on line-by-line measurement of RV shifts \citep{dumusqueMeasuringPreciseRadial2018,cretignierMeasuringPreciseRadial2020}, modelling of the observed spectrum as a linear combination of a time-invariant stellar spectrum and a time-variant telluric spectrum \citep[wobble,][]{bedellWOBBLEDatadrivenAnalysis2019}, or pairwise spectrum comparison through Gaussian process interpolation \citep[GRACE,][]{rajpaulRobustTemplatefreeApproach2020}.

In the next sections, we recast the template matching approach within a semi-Bayesian framework, and then evaluate the performance of template-based RV extraction methodologies when applied to ESPRESSO data. In particular, in Sect. \ref{Sec:data_processing} we discuss how the spectral data is processed, the stellar template is created and the telluric features are removed. Afterwards, in Sect.  \ref{Sec:chi2_method} we re-visit the classical template matching algorithm, and then in Sect. \ref{Sec:Bayes_method} we discuss the working principles of our semi-Bayesian template matching approach, as well as our strategy to efficiently characterize the posterior distribution of the model. In Sect. \ref{Sec:results} we evaluate the performance of our template matching algorithm using data from i) 22 HARPS observations of Barnard's star, to validate our template-based methodologies against the results of other template matching pipelines, when applied to a common set of spectra; and ii) 1046 ESPRESSO observations of 33 M-, K- and G-type stars, which will be the main focus of this paper. To the ESPRESSO dataset we will apply the classical template matching approach, the CCF of ESO's official pipeline and our semi-Bayesian approach, allowing for the comparison of RV scatter and uncertainties, as well as the estimation of an upper bound for RV precision. We refrain from studying the impact that stellar activity has on \texttt{S-BART} derived RVs, as our ESPRESSO sample mainly contains quiet stars. Further, such endeavor would translate into a large scale modelling effort that lies outside the scope of the current paper. Lastly, in Sect. \ref{Sec:limitations} we discuss some limitations of the developed methodology and present some possible improvements.

\section{Model preparation} \label{Sec:data_processing}
\begin{figure}[ht]
\resizebox{\hsize}{!}{\includegraphics{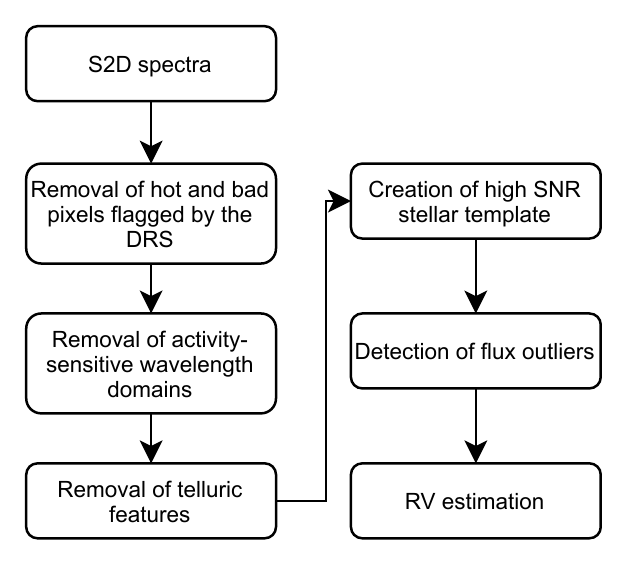}}
\caption{Workflow of the processing stage that we apply before the RV estimation.}
\label{Fig:preprocess_order}
\end{figure}

In this Section we discuss the stages of data processing, common to all instruments supported by our algorithm, that must be applied before estimating RVs from spectra. We also discuss the creation of the stellar template and removal of telluric features through the usage of synthetic spectra of Earth's atmosphere. Figure \ref{Fig:preprocess_order} shows the order in which the different procedures are applied.

\subsection{Pre-processing data} \label{Sec:pre-processing}
The extraction of the spectral orders from the image and necessary calibrations and corrections are handled by the official Data Reduction Software, DRS, of the respective instruments. Regions around spectral lines that are typically used as activity indicators or are clearly identifiable as emission features are removed from the spectra.

\begin{table}[ht]
\caption{Central wavelength (measured in air) and size of the spectral regions removed from the spectra.}              
\label{Tab:act_ind}      
\centering                                      
\begin{tabular}{c c c c}          
\hline\hline                        
Line & Wavelength (Air) [$\AA$]&Window [$\AA$] & Reference \\    
\hline                                   
CaK         &   3933.66          &   0.6        & 1 \\
CaH         &   3968.47          &   0.4        & 1 \\
$H\epsilon$ & 3970.075           &  0.6         & 3 \\
$H\delta$   & 4101.734           &  1.4         & 4 \\
$H\gamma$   & 4340.472           &  2.0         & 5 \\
$H\beta$    & 4861.35            &  1.8         & 6 \\
Na I D      &  5889.96           &   1.4        & 1 \\
Na I D      &  5895.93           &   0.9        & 1 \\
$H\alpha$   & 6562.808           &   2.0        & 2 \\
CaI         & 6572.795           &   1.8        & 2 \\
\hline                                             
\end{tabular}
\tablebib{
(1)~\cite{robertsonProximaCentauriBenchmark2016}; (2)~\cite{kuersterLowlevelRadialVelocity2003}; (3) Balmer series (n = 7 -> 2); (4) Balmer series (n = 6 -> 2); (5) Balmer series (n = 5 -> 2); (6) Balmer series (n = 4 -> 2), \cite{floresDiscoveryActivityCycle2016};
}
\end{table}

In Table \ref{Tab:act_ind} we identify the central wavelengths and the size of the spectral region that is removed around each feature. The chosen windows have been verified with the spectra of M-type stars. We also remove bad or hot pixels that are flagged by the  instrument's official pipeline, as well as those that have null data. If, when considering the mentioned effects, we remove more than 75\% of an order in a stellar spectrum we do not consider the order when estimating RVs.

\subsection{Telluric template} \label{Sec:telluric_template}

\begin{figure*}[tp]
\centering
\includegraphics[width=17cm]{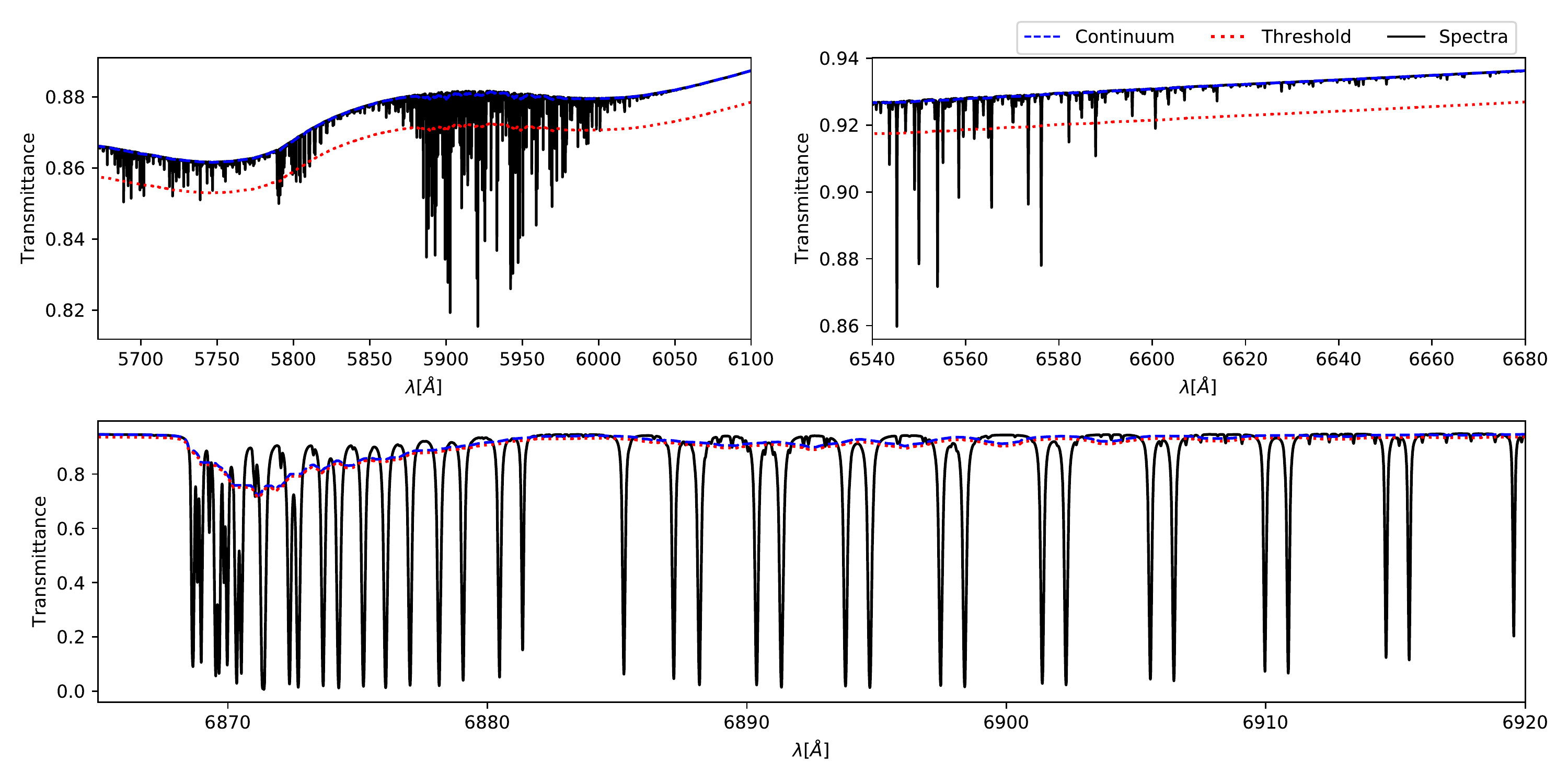}
\caption{Comparison, for three spectral regions, between the telluric spectrum obtained from Tapas (black) with the continuum level obtained with a median filter (dashed blue line) and the telluric threshold (dotted red line) built from the continuum.}
\label{Fig:telluric_template}
\end{figure*}

Earth's atmosphere absorbs radiation, imprinting telluric absorption features in spectra acquired with ground-based spectrographs. The impact of this phenomenon strongly depends on the wavelength range and resolution of the spectrograph, airmass of the observations, water vapor content and weather conditions \citep[e.g.][]{figueiraComparingRadialVelocities2012,cunhaImpactMicrotelluricLines2014}. If not corrected it can lead to biased and less precise RV estimates. Even shallow telluric lines, or micro-telluric lines, can induce a significant bias, about 10-20  ${\rm cm}\ {\rm s}^{-1}$ \citep{cunhaImpactMicrotelluricLines2014}, on par with, or larger than the signal produced by an Earth-like world around a solar-type star.

The identification and removal of telluric features from stellar spectra is thus essential for the estimation of accurate and precise RVs. For this purpose we use a synthetic spectrum of Earth's transmittance, with a resolution equal to that from the instrument mode, obtained through the Tapas \citep{bertauxTAPAS2014} web-interface\footnote{\url{http://cds-espri.ipsl.fr/tapas/}}. We start by estimating the continuum level of the transmittance spectrum through a rolling median filter which spans over 1000 points (though near the edges we reduce the window to 50 points to minimize numerical artifacts due to the choice of the filter's boundary conditions). We flag, as wavelength domains affected by tellurics, those where the transmission is lower than a given threshold - by default 99\% of the continuum level. Figure \ref{Fig:telluric_template} shows the behaviour of the rolling median filter. In regions with shallower telluric features the continuum estimation is not affected. However, that is no longer the case in regions where there is a larger presence of deeper features (bottom panel). Despite this, the chosen threshold is still enough to properly identify the telluric features, as seen in the bottom pannel of the Figure. It is important to refer that this choice of threshold is not able to detect shallower telluric features, as seen in the upper pannels of the Figure. However, a more restrictive threshold would result in a rejection of larger spectral regions across the wavelength coverage of the instrument. Thus, we attempt to maximize the spectral coverage whilst still removing the deeper telluric features.

We must also take into account the RV component introduced by Earth's motion around the barycenter of the Solar System (and as such nicknamed as Barycentric Earth Radial Velocity, or BERV). This motion introduces a Doppler shift in the observed spectrum that can be corrected by shifting the reference frame from Earth to an inertial one, the barycenter of the Solar System. This correction is incorporated, usually by default, in the spectrographs official  pipelines, as is the case of ESPRESSO, where the wavelength solution is shifted by the corresponding value. Since the telluric lines wavelengths are fixed on the detector on Earth's reference frame, their position relative to the stellar lines will change, and in the BERV-corrected spectra the telluric lines will appear as shifted by -BERV. To take into account this relative movement, we discard a wavelength domain around each feature corresponding to the maximum BERV ($\sim$ 30 ${\rm km}\ {\rm s}^{-1}$, obtained for stars along the direction of Earth's orbit).

\subsection{Stellar template} \label{Sec:stellar_template}
The stellar template is the most important component of our model, as it is assumed to be a high signal-to-noise model spectrum that represents very accurately the stellar spectrum, which is assumed to be immutable. Any observed spectrum is assumed to differ from this template only as a result of a Doppler-shift induced by the stellar RV. This high signal-to-noise template is built by combining the information of multiple observations of the same star.

\subsubsection{Building the template} \label{Sec:build_template}

The stellar template is built in an order-by-order basis, so that it can accurately represent the stellar spectra. Its construction starts with the choice of the reference frame for the template, i.e. the wavelength solution henceforth associated with it. For this purpose, we use the BERV-corrected observation with the smallest uncertainty in the RV estimated by the ESO pipeline (through the CCF method).

\begin{figure}[!ht]
\resizebox{\hsize}{!}{\includegraphics{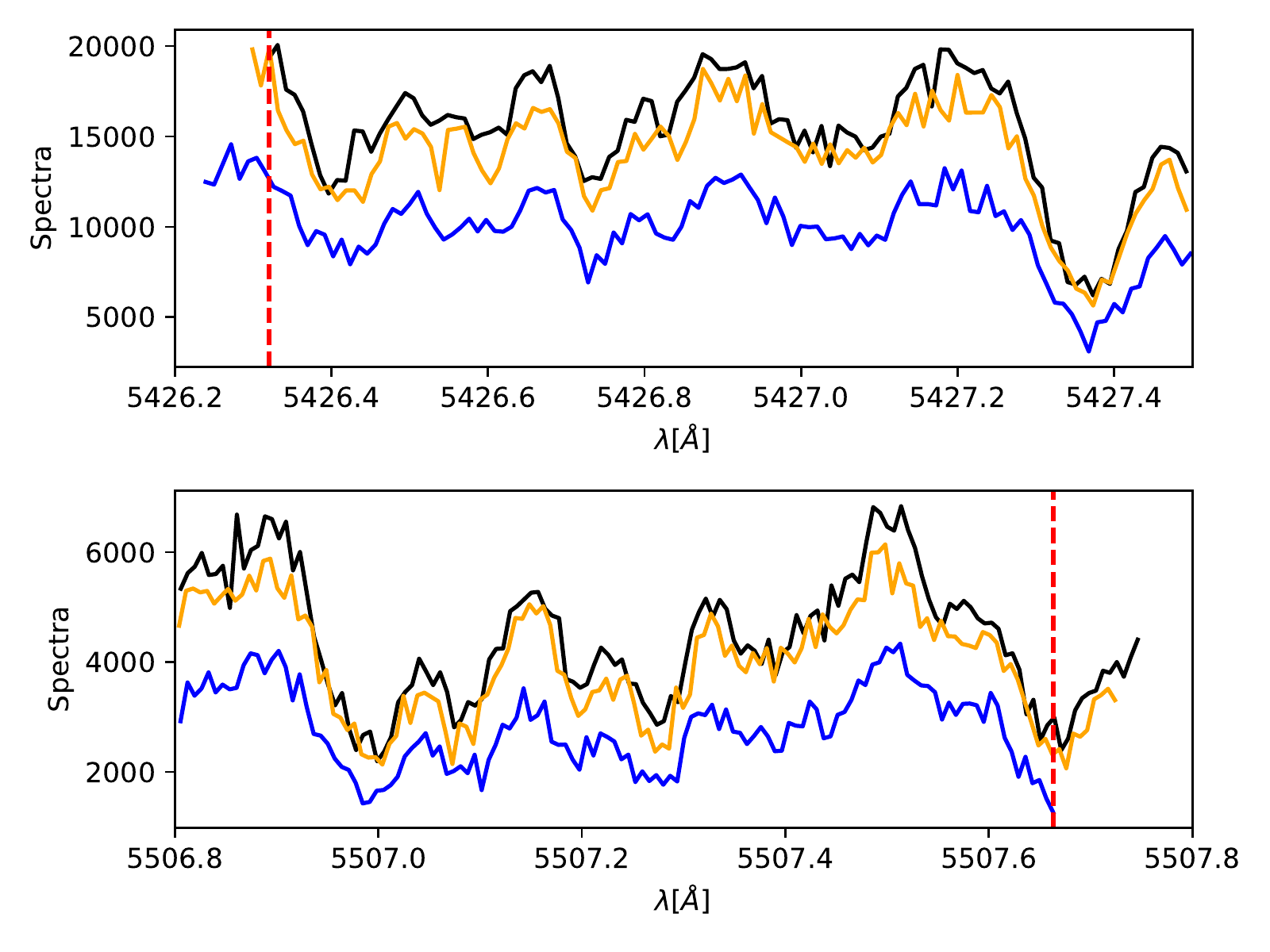}}
\caption{BERV-corrected spectra of different observations at the start (top) and end (bottom) of a spectral order. In the top panel the dashed red line represents the minimum wavelength so that all (of the presented) orders have data at the start of the order. In the bottom panel it represents the last wavelength at which all orders have data.}
\label{Fig:order_start_end}
\end{figure}

We decided to place our stellar template in a rest frame, i.e. at a RV of zero. To do so, we remove from the template's wavelengths the contribution of the stellar RV, either estimated beforehand through the CCF approach or a previous iteration of our template matching procedure. The next step is to remove, from all observations, the contribution of their own stellar RVs. As the wavelengths of the spectra will not be an exact match with those from the template, we have to interpolate them to a common wavelength grid - the one from the template. For this purpose we apply a cubic splines algorithm \citep[see Section 3.3 of][]{pressNumericalRecipesArt1992}. Due to the BERV, the stellar spectrum will shift on the CCD, and thus different spectra will have different starting and ending wavelengths in each spectral order (see Fig. \ref{Fig:order_start_end}). To avoid different SNR within the same order of the template  we select, for each order, the wavelengths common to all spectra. Finally, we compute the mean of the fluxes in order to build a high SNR stellar template. We calculate the mean in order to keep the count level at a physically meaningful value and avoid possible numerical issues further ahead.

Lastly, we must also consider the presence of telluric features in the data. Even though the wavelength domains affected by the deeper features can be removed with the methodology discussed in Sect. \ref{Sec:telluric_template}, micro-tellurics will not be identified by our mask and, consequently, still be present in the individual observations. As seen in \citet{cunhaImpactMicrotelluricLines2014} they can have a considerable impact in the accuracy and precision of the estimated RVs, in particular when obtained from data acquired with instruments as stable as ESPRESSO. By constructing the template from a large number of spectra, obtained at different periods of the year, i.e. by having a wide BERV range, their effects in the spectral template can, in principle, be minimized by averaging them out. However, as this condition is not always met, we mitigate their impact by using in the building of the spectral template only observations whose associated airmass is smaller than 1.5\footnote{This value was selected as the default one, but it can quickly be changed to accommodate the observing conditions of the available observations.} as the depth of the telluric features increases with airmass \citep{figueiraEvaluatingStabilityAtmospheric2010}. This choice allows to strike a balance between i) the number of observations that are discarded due to high micro-telluric contaminations and ii) the number of observations that can be used in the construction of the template. Furthermore, at higher airmasses the correction of the atmospheric dispersion is not as efficient as it is for lower airmasses \citep{wehbeAtmosphericDispersionCorrection2019}. Thus, this selection is an attempt to select a set of homogeneous spectra to be used in the construction of the stellar template.

\subsubsection{Estimation of uncertainty in the stellar template} \label{Sec:template_uncerts}

The stellar template will be affected by some uncertainty, given that it is built as a mean of $N$ spectra, all affected by flux measurement uncertainties. Within this sub-Section we start by discussing the calculation of the uncertainties associated with the stellar template, following with a comparison with the uncertainties in both low- and high-SNR observations. Lastly, we touch upon the computational trade-offs that must be made to ensure performance of the algorithm.

If each spectrum had the same SNR, then we would expect the SNR of the template to be approximately equal to $\sqrt{N}$ times the SNR of each observation. Under this assumption, we would need 100 observations to achieve a mean uncertainty (standard deviation) per flux one order of magnitude smaller than that associated with any single observation. Since many targets will have fewer observations, we decided to propagate the uncertainties associated with the template towards the final RV estimate, as discussed in Sects. \ref{Sec:chi2_method} and \ref{Sec:Bayes_method}. For this purpose, we have to take into account that both the spectral data and the template are interpolated with cubic splines in two different stages: during the creation of the stellar template and in the RV extraction procedure. In Appendix \ref{App:error_propagation} we describe the analytical uncertainty propagation through the cubic spline interpolation algorithm.

We studied the characteristics of the uncertainty in the template by selecting the available observations of an M4 star ($N_{spectra}$=21) from the sample used in Sect. \ref{Sec:ESPRESSO_rms_comparison}, allowing to assess the need of accounting for them during the RV estimation. The chosen observations were made after ESPRESSO's fiber link upgrade in June 2019 \citep{pepeESPRESSOVLTOnsky2021} and we selected data from the 100th spectral order (central wavelength of 541 nm). In order to evaluate the impact of the number of spectra used to construct the template we selected two sets of observations:  those with an airmass below 1.2 (N = 8); and those with an airmass below 1.5 (N=13). After creating the stellar templates we align them with each observation and interpolate the template's flux to the wavelength solution of the observation, also propagating the uncertainties in the template. Then, we compare them against the ones associated with the spectra, as computed by the ESO pipeline. A direct comparison is not possible, as observations with lower flux values also have lower photon noise and, consequently, smaller flux uncertainties. Instead, we compute the SNR ratios of the template and spectra, for each pixel of the order.

\begin{figure}[!ht]
\resizebox{\hsize}{!}{\includegraphics{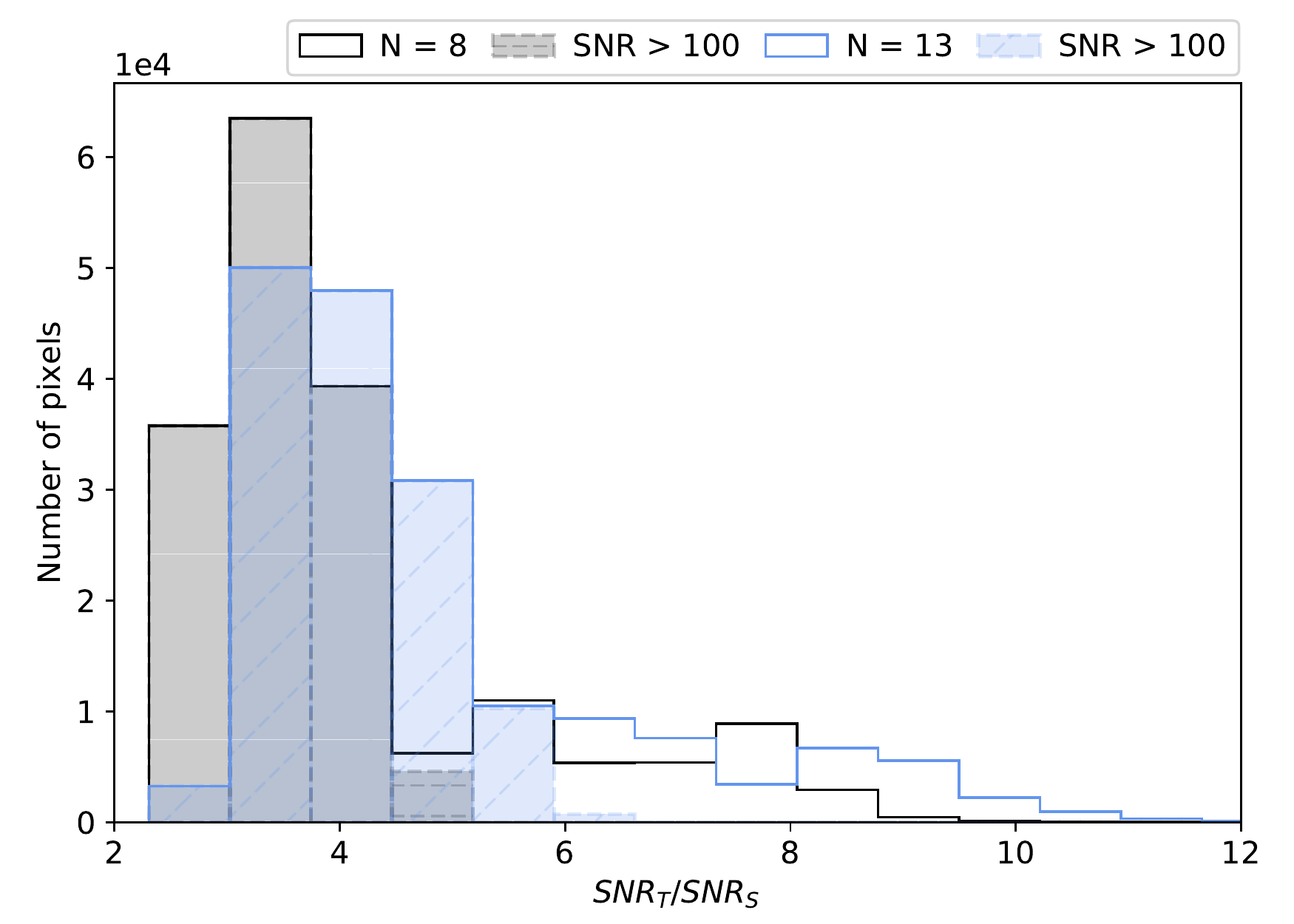}}
\caption{Histogram of the SNR ratio, for the 100th order, of the template and each of the 21 individual observations of an M4 star, obtained after the fiber link upgrade of ESPRESSO in mid 2019. In order to compare the impact of the number of observations used to construct the template in its associated uncertainty, one template was built with 8 observations (black curve) and the other with 13 observations (blue curve), selected by the airmass at the start of each observation. For all observations we also highlight by filling the bins with the corresponding colour, the comparison with observations with a SNR of at least 100 for the selected spectral order.}
\label{Fig:template_uncert_evol}
\end{figure}

Figure \ref{Fig:template_uncert_evol} shows that an increase in the number of observations used to construct the template leads to an increase in its SNR, when compared against the individual observations, as one would expect. The comparison for the lower SNR spectra (non-filled bins) reveals that the two stellar templates have uncertainties close to one order of magnitude smaller than the ones found on the observations. Unfortunately, that is not the case for the higher SNR observations, i.e. those with an SNR in the 100th order of at least 100. For a more detailed analysis, Table \ref{Tab:template_uncert_statistics} shows the comparison of each template against three different sets of observations: with all observations; with only the observations used in the construction of the template; with observations whose SNR is at least 100. We find that in all cases the SNR median value is larger than $\sqrt{N}$, a difference explained by the fact that the selected observations all have different SNRs. From this analysis we see that the SNR of the template is not one order of magnitude larger than the one from the observations, confirming the need of accounting for those uncertainties.

\begin{table}[ht]
\caption{Analysis of the SNR ratio between each of the two templates and three different subsets of the available observations.}
\label{Tab:template_uncert_statistics}
\centering
\begin{tabular}{c c c c c}
\hline\hline
Airmass & N & $\sqrt{N}$ & Observations & SNR ratio \tablefootmark{a} \\ \hline
\multirow{3}{*}{$\leq$ 1.2} &\multirow{3}{*}{8} & \multirow{3}{*}{2.8} & All\tablefootmark{b} & 3.7 $\pm$ 2.1 \\
& & & Template \tablefootmark{c}& 3.4 $\pm$ 0.5 \\
& & &  SNR $\geq$ 100 \tablefootmark{d}&  3.3 $\pm$ 0.5 \\ \hline
\multirow{3}{*}{$\leq$ 1.5}  &\multirow{3}{*}{13} & \multirow{3}{*}{3.6}   & All\tablefootmark{b} & 4.4 $\pm$ 1.9 \\
& & & Template\tablefootmark{c} & 4.1 $\pm$ 0.9 \\
& & & SNR $\geq$ 100 \tablefootmark{d} & 3.9 $\pm$ 0.6 \\
\hline
\end{tabular}
\tablefoot{
 \tablefoottext{a}{The values and the associated uncertainty represent the median and standard deviation of the SNR ratios;}
 \tablefoottext{b}{Compares the two templates against all available observations from the two sets;}
 \tablefoottext{c}{Comparison against only those that were used to construct the corresponding stellar template;}
 \tablefoottext{d}{Comparison against all observations that have a SNR, in the 100th order, higher than 100.}
}
\end{table}

The main problem with our uncertainty propagation procedure, described in Appendix \ref{App:error_propagation}, lies in its computational efficiency, as it requires the inversion of large matrices. Whilst it is feasible to use it for the calculation of uncertainties during the creation of the stellar template, it is a computational bottleneck in the interpolation of the stellar template for each tentative RV shift that is tested during the RV extraction procedure. In an attempt to mitigate this problem, we decided to evaluate if we could estimate the template flux uncertainties by interpolation, instead of applying the analytical uncertainty propagation procedure.

This approximation was tested through the interpolation of the two templates previously used in Fig. \ref{Fig:template_uncert_evol}, and then comparing the flux uncertainties obtained through both methods. We found, for the two templates, that the interpolation of flux uncertainties (standard deviations) results in their overestimation by a factor of $1.07\pm 0.05$ across the 21 observations. Even though there is a slight increase in the template uncertainty, we do not deem it to be problematic, especially when also considering the high-SNR of the template itself, when compared against the individual observations. With this in mind we propagate, during the RV extraction, the uncertainties in the stellar template by interpolating them to the desired wavelength solution.

\subsubsection{Removal of outliers in the spectra} \label{Sec:outlier_temoval}

\begin{figure}[!ht]
\resizebox{\hsize}{!}{\includegraphics{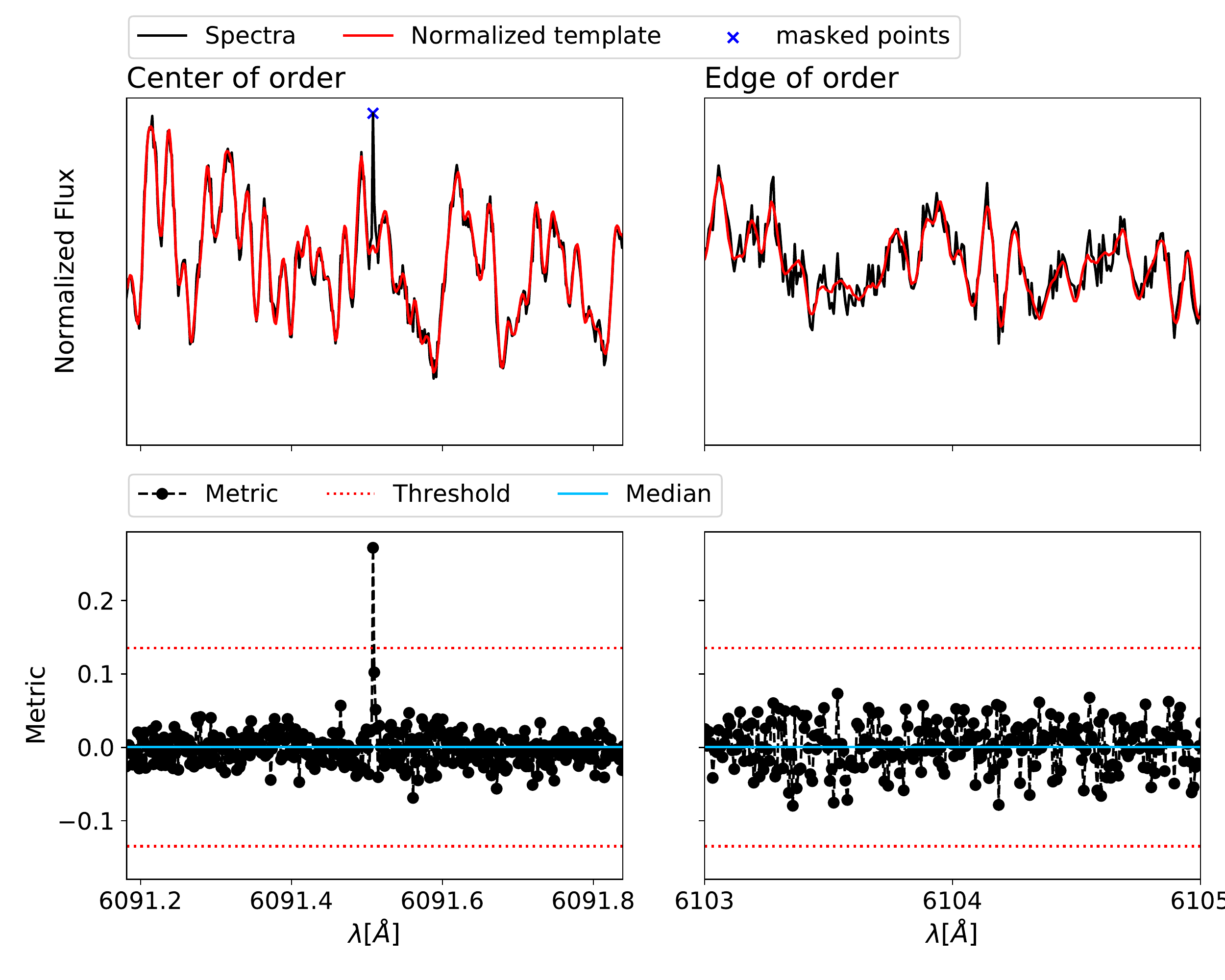}}
\caption{Outlier removal routine for two regions of the same order, with respect to an observation for which outlier identification achieved convergence after the first iteration. For representation purposes we normalized the flux in the center and the edge of the order. \textbf{Top row:} Comparison between the stellar template (red line) and the spectra (black line). The blue crosses represent the points that were flagged by the method. \textbf{Bottom row:} Differences between the template and spectra (black points). The blue line is the median value, whilst the dotted red lines represent a 6$\sigma$ difference from it.}
\label{Fig:template_outliers}
\end{figure}

Even though the stellar template is generally a good match to the stellar spectrum associated with any given observation, there are some regions where such assumption does not hold, e.g. as seen in the top row of Fig. \ref{Fig:template_outliers}, thus raising the need to remove so-called flux outliers before starting the RV estimation procedure. It is important to note that any point that was discarded in Sect. \ref{Sec:pre-processing} will be ignored during the search for outliers.

We start by aligning the stellar template and a given spectrum using the initial guess for the associated RV, either estimated through the CCF method or a previous application of template matching. Then we adjust both continuum levels by fitting a first degree polynomial, with slope $m$ and intercept $b$, to the ratio between the spectrum and template. Finally, we compute the logarithm of the ratio between spectrum and template and use it as a metric to flag mismatch regions:

\begin{equation}
metric_{\lambda_i} = \log\left( \frac{S_{\lambda_i}}{p(m,b)_{\lambda_i}T_{\lambda_i}}\right)
\end{equation}

\noindent where $S_{\lambda_i}$ is the flux of the stellar spectrum, $p(m,b)_{\lambda_i}$ the first degree polynomial and $T_{\lambda_i}$ the interpolated stellar template, all evaluated at wavelength ${\lambda_i}$. We use the logarithm, instead of the ratio, in order to mitigate the larger differences that exist for lower SNR regions closer to the edges of the spectral orders. This metric ensures performance for a large dynamic range of fluxes as seen inside a single order. Further, as the spectra are noisy and we are mostly interested in finding large flux differences, we allowed for a large tolerance in the identification of outliers. We consider all points whose metric is more than 6$\sigma$\footnote{We use $\sigma$ to refer to the standard deviation of the metric across the entire order.} away from the median metric (of the entire order) to be outliers. Using lower thresholds would result in excessive flagging in both edges of the order, as the differences (in absolute values) between spectra and template are larger.

In Fig. \ref{Fig:template_outliers} we can see an application of the algorithm to two spectral chunks. Starting in the left panel, we find minimal (relative) differences between the spectrum and the normalized template, with the exception of the clear outlier that is flagged by our routine. On the right panel we see that the spectrum is noisier and, consequently, the match between spectrum and template is worse. Nonetheless, our algorithm does not flag any point, as there are no clear outliers.

\section{Classical template matching} \label{Sec:chi2_method}

\begin{figure*}[tp]
\centering
\includegraphics[width=17cm]{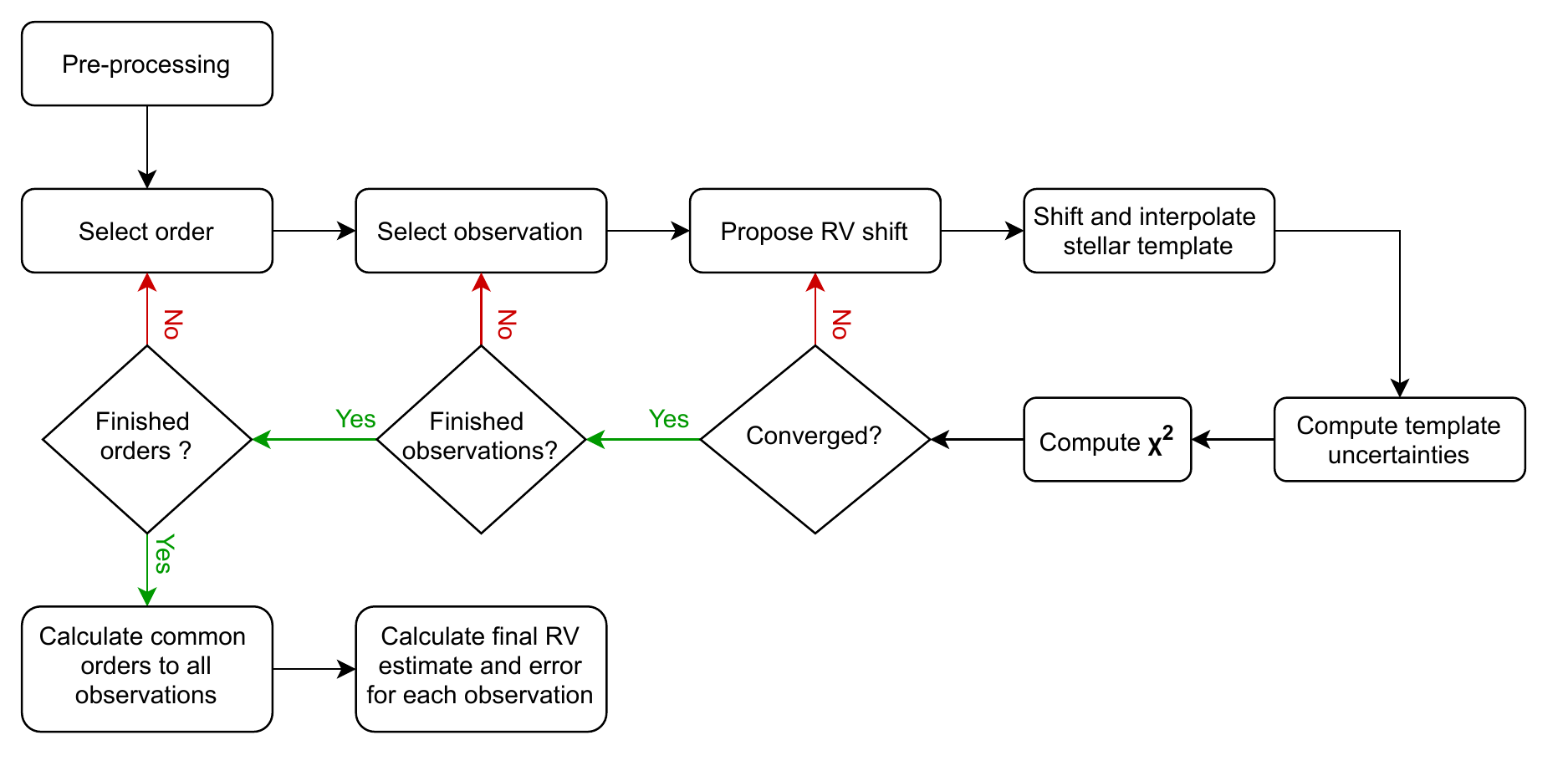}
\caption{Schematic of the Classical template matching RV estimation procedure, including the computation of the uncertainties in the stellar template. We iterate over orders at the highest level, not observations, in order to optimize the computational efficiency of the method. }
\label{Fig:gj699_scheme}
\end{figure*}

In order to benchmark our methodology to estimate radial velocities we first implemented a template matching approach similar to those used to build the \texttt{HARPS-TERRA} \citep{anglada_escude_HARPS_TERRA_2012} and \texttt{SERVAL} \citep{zechmeisterSpectrumRadialVelocity2018} pipelines. In Fig. \ref{Fig:gj699_scheme} we provide an high-level schematic of the RV estimation procedure, whose boxes are described with more detail within this Section.

Radial velocities are determined individually for each order \textit{i}, through least squares minimization:

\begin{equation} \label{Eq:chi_square}
\chi^2 = \sum_{i=1}^{N_{pixels}} \frac{1}{\boldsymbol{\sigma_S^2 + \sigma_T^2} } * [S_{\lambda_i} - p(m,b)_{\lambda_i}T_{\lambda_i}]^2
\end{equation}

\noindent where $S_{\lambda_i}$ is the flux of the stellar spectrum, $\boldsymbol{\sigma_S}$ its associated (1$\sigma$) uncertainty, $p(m,b)_{\lambda_i}$ the first degree polynomial mentioned in Sect. \ref{Sec:outlier_temoval}, $T_{\lambda_i}$ the stellar template and $\boldsymbol{\sigma_T}$ its (1$\sigma$) uncertainty, all evaluated at wavelength $\lambda_i$.

The $\chi^2$ minimization is performed with \texttt{scipy}'s \citep{virtanenSciPyFundamentalAlgorithms2020} implementation of the Brent method, inside a window with a default size of 200 ${\rm m}\,{\rm s}^{-1}$ centered around the previous RV estimate. The stellar template is Doppler-shifted by each tentative RV and interpolated, with a cubic spline algorithm, to the wavelength solution of the spectra. Similarly to the creation of the stellar template, Sect. \ref{Sec:template_uncerts}, the uncertainties of the stellar template are also interpolated to the new wavelength solution. After the minimizer converges, we use the proposed RV value and two adjacent ones, separated by the RV interval $\pm\Delta RV$, to numerically fit a parabola to the curve, similarly to \citet{zechmeisterSpectrumRadialVelocity2018}. By default, we assume $\Delta RV$ to be 10 ${\rm cm}\,{\rm s}^{-1}$ for ESPRESSO and 50 ${\rm cm}\,{\rm s}^{-1}$ for HARPS. For further details with respect to this fit we refer to Sect. 10.2 of \citet{pressNumericalRecipesArt1992}. The fitted parabola is then used to correct the RV estimate and calculate its uncertainty

\begin{align} \label{Eq:parabola_min_point}
\begin{split}
RV_{order} &= RV_{min} - \frac{\Delta RV}{2} \frac{\chi^2_{m+1} - \chi^2_{m-1}}{\chi^2_{m+1} + \chi^2_{m-1} - 2\chi^2_m} \\
\boldsymbol{\sigma_{RV}^2} &= \frac{2\Delta RV^2}{\chi^2_{m-1} - 2\chi^2_m + \chi^2_{m+1}} \\
\end{split}
\end{align}

\noindent where $RV_{order}$ is the final RV estimate, $\boldsymbol{ \sigma_{RV} }$ is the measurement (1$\sigma$) uncertainty, $RV_{min}$ is the RV value that minimizes Eq. \ref{Eq:chi_square}, while $m-1$ and $m+1$ identify the two RV values selected at a distance $\Delta RV$ from $RV_{min}$.

If the minimizer proposes a value near the edges of the search window, there is no guarantee that the proposal is the true minimum of the $\chi^2$ curve. Thus, whenever the result of the Brent minimization lies within 5$\Delta RV$ of either edge of the search window we define a new one, centered at the edge of the interval, with a size of 30 ${\rm m}\,{\rm s}^{-1}$ and re-start the minimization procedure. As we do not expect such large differences between the CCF RVs and those obtained through a template matching method, we discard the entire order if we, once again, do not find a minimum RV that meets the aforementioned criteria within the new search window.

\begin{figure}[!ht]
\resizebox{\hsize}{!}{\includegraphics{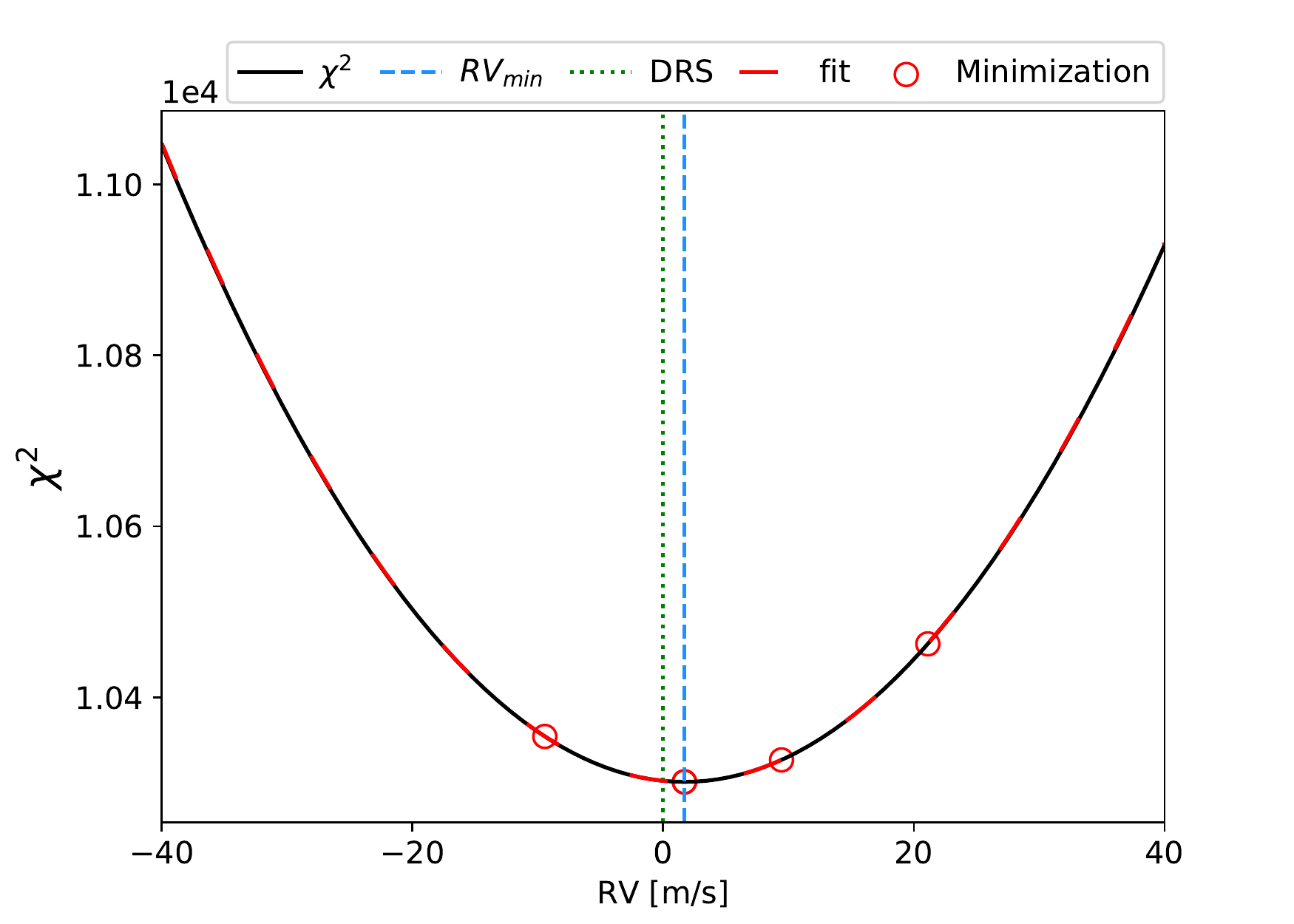}}
\caption{The $\chi^2$ curve for the 100-th spectral order of an ESPRESSO observation of GJ699 (black line), sampled with a step of 10 ${\rm cm}\,{\rm s}^{-1}$ inside the 200 ${\rm m}\,{\rm s}^{-1}$ search window centered on the ESPRESSO DRS estimate. The parabolic fit is shown as the red line. The red circles represent the RV estimates provided by the Brent method before meeting the convergence criteria. The dashed blue line is the final RV, calculated with Eq. \ref{Eq:parabola_min_point}, and the dotted green line the ESPRESSO DRS estimate for the given observation.}
\label{Fig:gj699_chicurve}
\end{figure}

In Fig. \ref{Fig:gj699_chicurve} we show the comparison of the CCF-based ESPRESSO pipeline (DRS) RV estimate and that obtained through Equation \ref{Eq:parabola_min_point} for one spectral order. There is a very good match between the parabolic fit and the full $\chi^2$ curve. Furthermore, the advantages of using the Brent method for the minimization are evident: it can quickly achieve convergence, greatly increasing the computational efficiency of the routine (e.g. in the case shown in Fig. \ref{Fig:gj699_chicurve} we only have to sample Eq. \ref{Eq:chi_square} four times before achieving convergence for the spectral order).

After estimating a RV with respect to each individual order, we combine them through a weighted mean
with inverse variance weights \citep{schmellingAveragingCorrelatedData1995} in order to calculate the final RV estimate (${\rm v}$) and uncertainty ($\boldsymbol{ \sigma_{\rm v} }$)  for the entire observation:
\begin{align} \label{Eq:final_rvs}
\begin{split}
{\rm v} &= \frac{\sum \sigma^{-2}_{{\rm v}_i} {\rm v}_i}{\sum \sigma^{-2}_{{\rm v}_i}} \\[1ex]
\boldsymbol{  \sigma_{\rm v} } &= \sqrt{\frac{1}{\sum 1/\sigma^{2}_{{\rm v}_i}}}
\end{split}
\end{align}

\noindent where ${\rm v}_i$ is the RV of order $i$ and $\boldsymbol{ \sigma_{{\rm v}_i} }$ its associated ($1\sigma$) uncertainty, while $N$ is the total number of orders for which we estimated an RV. The value of $\boldsymbol{ \sigma_{{\rm v}_i} }$ is estimated whilst ignoring any correlations that might exist between the assumed independent spectral orders.

As we have seen in this Section, and in Sect. \ref{Sec:pre-processing}, some spectral orders may be discarded for some of the observations. In this case, different wavelength domains would be used in the estimation of the final RVs of the different observations. This could introduce additional RV variability within the considered set of observations, due to possible differences in the spectral information contents and systematic effects in each spectral order. In order to avoid this additional source of RV variability we retain only the wavelength domain that is common (i.e. not discarded) to all observations. The drawback to this approach is that the inclusion of a single observation can cause the removal of a large number of orders. In this case there would be a smaller loss of information if the entire observation was discarded instead. We currently have no way of evaluating this trade-off other than doing a manual inspection of the number of orders removed per observation, and then determining how all the RV final estimates and its uncertainties change in the case one or more observations are discarded from the full RV estimation procedure.

\section{Semi-Bayesian template matching} \label{Sec:Bayes_method}

The current approaches to template matching assume Doppler-shifts associated with the different spectral orders are independently generated. However, this is clearly not the case for the stellar RV component induced by orbiting bodies, like planets or companion stars. In fact, such shifts are achromatic, i.e. they are independent of the wavelengths at which they are measured. Relative RV estimates, as those obtained through template matching, are primarily used to detect orbiting bodies and characterize their masses and orbits. Thus, consistency then suggests that one should use a single RV shift to describe simultaneously the differences for all orders between a given spectrum and the stellar template. Any effects that may hinder the correct estimate of this single RV shift, like those of instrumental origin or due to stellar activity, should be dealt with explicitly, either through modelling or exclusion of affected spectral data.

\begin{figure*}[tp]
\centering
\includegraphics[width=17cm]{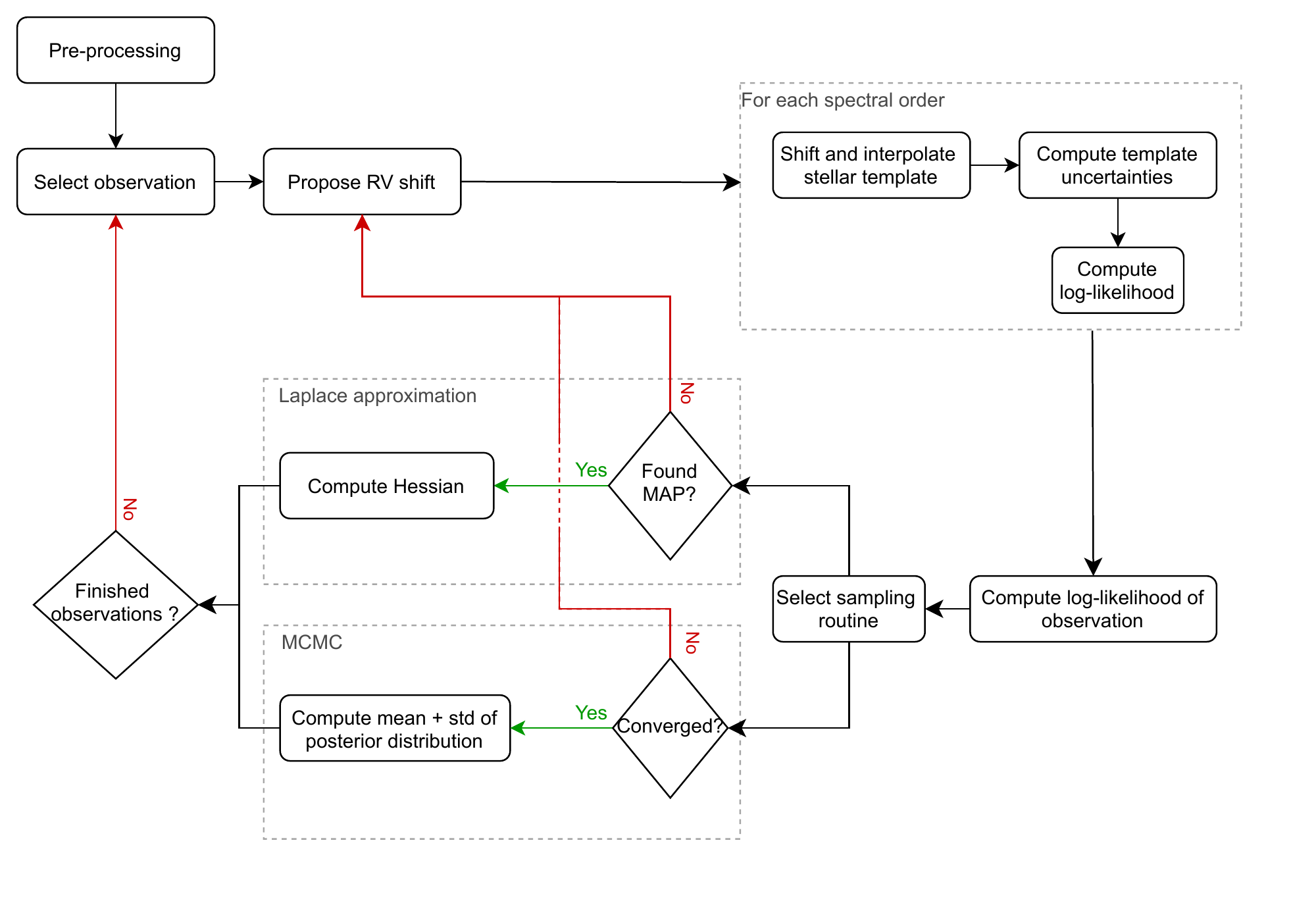}
\caption{Schematic of the semi-Bayesian approach, considering both an MCMC and Laplace approximation to characterize the posterior distribution.}
\label{Fig:bayesian_scheme}
\end{figure*}

The casting of RV estimation through template matching into a Bayesian statistical framework allows for consistent and straightforward characterization of the RV (posterior) probability for any observation, including  marginalization with respect to the parameters of the first-degree polynomials that are used to adjust the continuum level of spectra and template. However, within a Bayesian framework all aspects of the model considered need to be specified prior to the actual data analysis, i.e. the information contained in the data cannot be used twice, for building the model (prior specification) and also for comparison with its predictions (through the likelihood). Unfortunately, the later takes place in the context of template matching, because the available spectra are used to specify the template (model building) as well as to estimate the RVs at the times of spectra acquisition (data analysis). This is the reason why we call semi-Bayesian to the template matching approach for RV estimation described schematically in Fig. \ref{Fig:bayesian_scheme} and with greater detail in this Section.

This approach has been implemented in a pipeline capable of processing HARPS and ESPRESSO data, which has been named S-BART: Semi-Bayesian Approach for RVs with Template-matching\footnote{Publicly available in \url{https://github.com/iastro-pt/sBART}}. It is important to note that the \texttt{S-BART} pipeline allows for the usage and configuration of all techniques that have been described in this manuscript, including the traditional template matching approach, i.e. the one introduced in Sect. \ref{Sec:chi2_method}.

\subsection{The RV model} \label{Sec:Bayes_RV_model}

We apply our RV model independently to each observation. It contains only one parameter of interest, the relative RV with respect to some reference frame, which is the one for which the template RV is zero. We thus wish to characterize the RV posterior probability distribution given each observed spectrum, which is proportional to the product of the RV prior probability distribution by the likelihood of the spectral data. We will assume a uninformative prior, taking the form of an uniform probability distribution. The likelihood of the full spectral data, conditioned on a given RV value, will be assumed to equal the product of the likelihoods of the fluxes measured for each pixel, i.e. the flux measurements for all pixels are considered to be independent. In practice, we first calculate the likelihood of the spectral data for each order, and then multiply these to obtain the full likelihood.

However, our RV model also contains so-called nuisance parameters that we have to marginalize over. The nuisance parameters are those involved in the template matching procedure described in Sect. \ref{Sec:stellar_template}:
\begin{itemize}
    \item The slope, $m$, and intercept, $b$, of the first degree polynomial that is used to adjust the continuum level of the template spectrum to that of each spectrum under analysis;

    \item  The flux associated with each pixel in the template, relative to the continuum.
\end{itemize}
 The latter are effectively latent variables, affected by some uncertainty, characterized in Sect. \ref{Sec:template_uncerts}. Each spectral order has its own set of independent nuisance parameters, thus the marginalization procedure can be applied order by order. We assume the joint prior distribution with respect to all model parameters to be separable, i.e. can be written as the product of prior distributions specifically associated with each parameter. Thus, in practice, the marginalization procedure involves the integration of the product of the prior distributions with respect to the nuisance parameters by the likelihood as a function of all parameters.

The likelihood of a given observed spectrum, $S$, conditioned on an assumed RV value, is thus given by

\begin{multline} \label{Eq:marginalization}
P(S | RV) = \prod\limits_{i=1}^{N_{\rm orders}}\int P(m_{i},b_{i})\prod\limits_{j=1}^{N_{i}}P(T_{\lambda_{i,j}})\times\\
\times P(S_{\lambda_{i,j}}|{\rm RV},m_{i},b_{i},T_{\lambda_{i,j}}) \,{\rm d}m\,{\rm d}b\,{\rm d}T_{\lambda_{i,j}}
\end{multline}

\noindent where $N_{\rm orders}$ is the total number of orders in the spectrum
that are not discarded as a result of the procedure discussed in Sect. \ref{Sec:chi2_method}, $N_{i}$ is the number of data points in order $i$ (usually smaller than the number of pixels in the order, due to the masking of spectral regions contamination by telluric lines and outlier removal), $T_{\lambda_{i,j}}$ and $S_{\lambda_{i,j}}$ are the fluxes associated with pixel $\lambda_{i,j}$ in the (interpolated) template and spectrum, respectively. We will assume that the prior probabilities, $P(m_{i},b_{i})$ and $P(T_{\lambda_{i,j}})$, are uninformative Gaussians. Given that the last probability is also a Gaussian, with an expected value that is a linear function of ${\lambda_{i,j}}$, i.e. $(b+m\lambda_{i,j})T_{\lambda_{i,j}}$, in the limit of infinite variance for the prior probabilities, the integral is equal to a Gaussian and

\begin{equation} \label{Eq:marginal_equation}
\begin{split}
\log P(S | RV) = \sum\limits_{i=1}^{N_{\rm orders}} \left(-\frac{1}{2}S_i^TK_i^{-1}S_i + \frac{1}{2}S_i^TCS_i -  \frac{1}{2}\log\ |K_i| - \right.\\\left. - \frac{1}{2}\log\ |A| - \frac{n_i-n_H}{2}\log\ 2\pi \vphantom{\int_1^2} \right)
\end{split}
\end{equation}

\noindent where $S_i$ is a vector with the flux measurements for all pixels in order $i$, A is equal to $H_{i}K_{i}^{-1}H_i^T$, C is given by $K_i^{-1}H_{i}^{T}A_{i}^{-1}H_{i}K_i^{-1}$, and $n_i$ is the number of data points in order $i$ \citep[for more details see Sect. 2.7 of][]{rasmussenGaussianProcessesML_2006}. The $2\times n_i$ matrix $H_i$ contains the values of the $n_H=2$ basis functions associated with the linear model $E[S_{\lambda_{i,j}}]=(b+m\lambda_{i,j})T_{\lambda_{i,j}}$ for each data point, i.e. $1$ and $\lambda_{i,j}$ with the associated parameters $bT_{\lambda_{i,j}}$ and $mT_{\lambda_{i,j}}$, respectively. Finally, the variance-covariance matrix $K_i$ is diagonal with entries $\sigma_{S_{\lambda_{i,j}}^2}+\sigma_{T_{\lambda_{i,j}}^2}$, where $\sigma_{S_{\lambda_{i,j}}}$ and $\sigma_{T_{\lambda_{i,j}}}$ are the standard deviations associated with the Gaussian probability distributions that describe the uncertainties, respectively, in the flux measurement and stellar template construction (including the cubic spline interpolation) processes.

The construction of the RV estimate is made through the RV posterior distribution, further discussed in Sect. \ref{Sec:rv_posterior_char}.  Its mean value and standard deviation will provide estimates of the RV value and uncertainty, respectively. We summarize the posterior with its mean value, as this minimizes the quadratic loss function associated with this estimation. The uncertainty in the RV estimates will thus account for all sources of noise, including noise in the observed spectra and stellar template.

\subsection{Characterization of the RV posterior distribution} \label{Sec:rv_posterior_char}

We used the \textit{emcee} package \citep{foreman-mackeyEmceeMCMCHammer2013} to characterize the RV posterior probability distribution associated with each observation through the \textit{Markov Chain Monte Carlo} (MCMC) methodology. We assessed MCMC convergence through several criteria that must be met simultaneously:

\begin{itemize}
    \item The chain length must be at least 50 times larger than the autocorrelation time ($\tau$);
    \item The value of $\tau$ cannot change more than 1\% after each iteration;
    \item The mean and standard deviation of the chains cannot change by more than 2 \% after each iteration.
\end{itemize}

Unfortunately, achieving convergence for a single ESPRESSO observation takes $\sim$10 minutes on a 24 cores@2.3GHz and 128 GB RAM server, making the method computationally expensive for stars with a large number of available observations.

\begin{figure}[!ht]
\resizebox{\hsize}{!}{\includegraphics{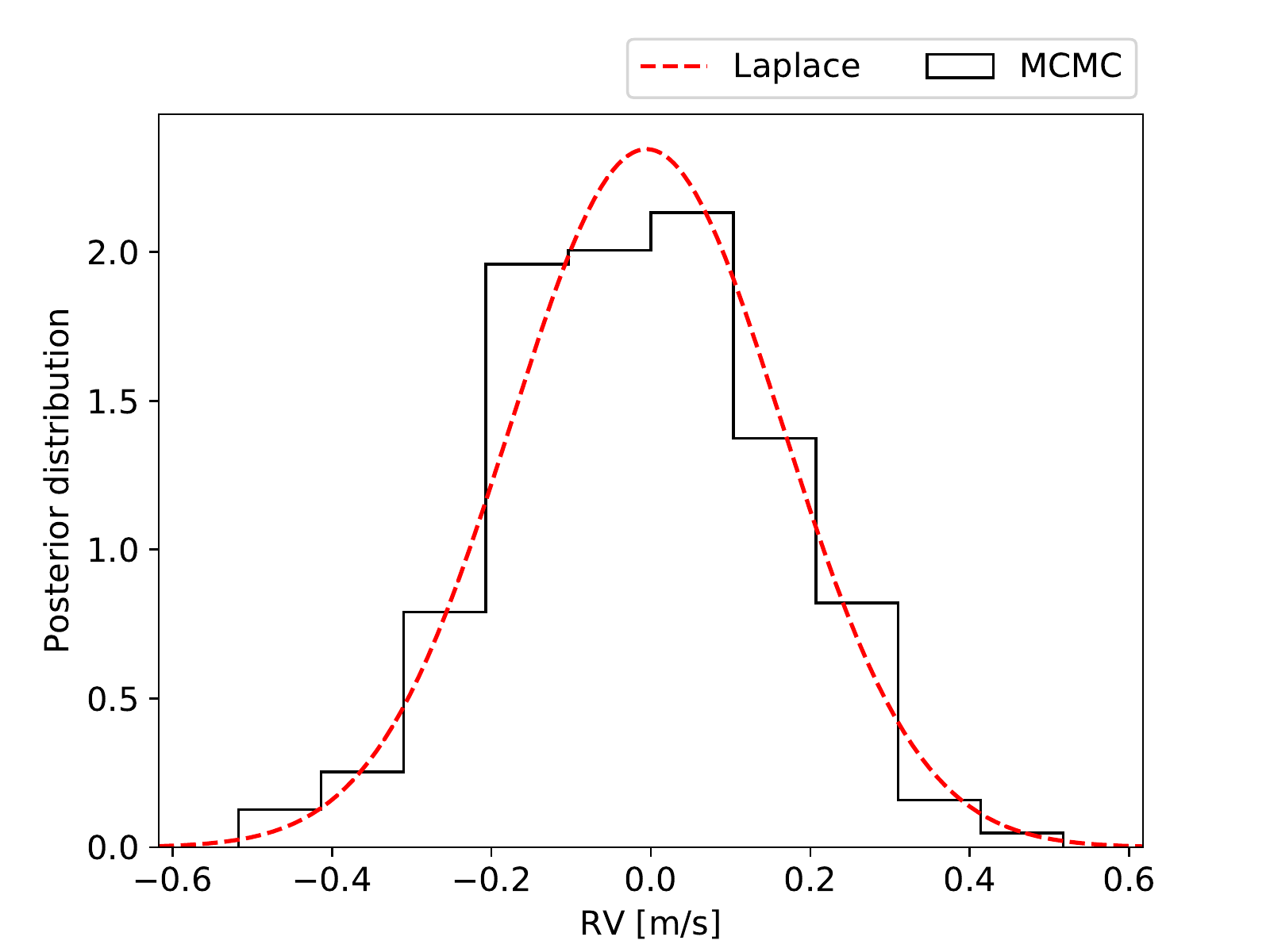}}
\caption{Comparison of the RV posterior distribution derived using the MCMC methodology (black curve) with the result of the Laplace approximation (dashed red line), for one observation of the M4 star considered in Sect. \ref{Sec:template_uncerts}.}
\label{Fig:mcmc_laplace_comp_single_obs}
\end{figure}

However, since the posterior distribution for the RV shift is approximately Gaussian (due to the large amount of information in the data), we can use the Laplace approximation to characterize it. By performing a second order Taylor expansion around the posterior distribution maximum we can approximate it with a Gaussian distribution centered at the posterior's mode (also known as MAP - maximum a posteriori) and variance equal to the inverse of the Hessian of the log-posterior evaluated at the mode \citep[e.g. see Sect. 3.4 of][]{rasmussenGaussianProcessesML_2006}, as shown in Fig. \ref{Fig:mcmc_laplace_comp_single_obs}. In practice, this approximation transforms the characterization of the posterior into a optimization problem - the minimization of the negative log likelihood (they are equivalent, since we assume a uniform prior for the RV). To solve it we apply, once again, \textit{scipy}'s implementation of the Brent method, reducing the computational time to $\sim30$ seconds per observation on the same machine.

\begin{figure}[ht]
\resizebox{\hsize}{!}{\includegraphics{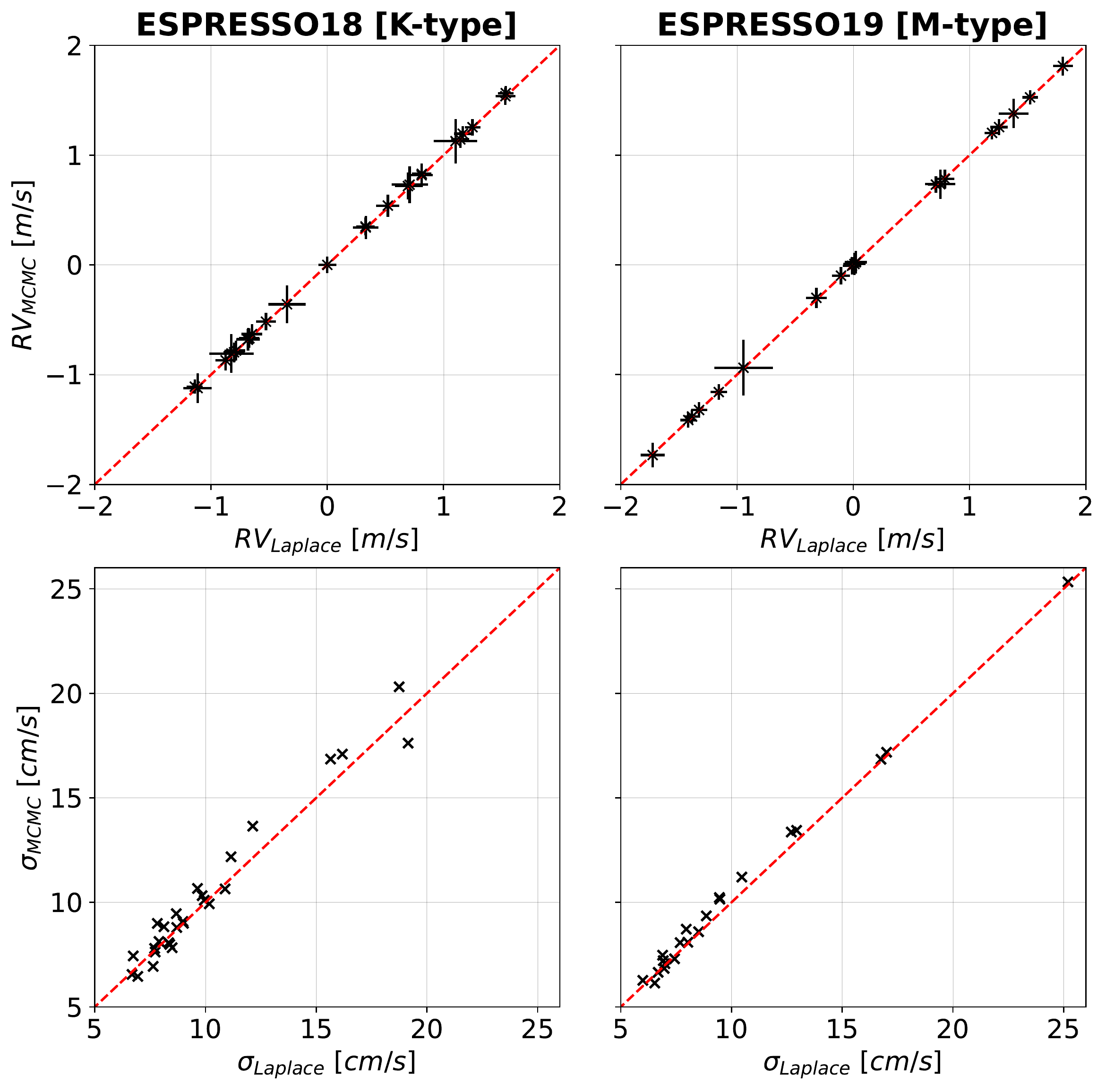}}
\caption{Differences between the RV posterior distribution as characterized through the MCMC methodology and the Laplace approximation, for 27 ESPRESSO18 observations of a K-type star (left column) and 21 ESPRESSO19 observations of a M-type star (right column). \textbf{Top pannel:} Comparison of the RV expected value, in  ${\rm m}\ {\rm s}^{-1}$.  We show, in each axis, the uncertainty associated with the corresponding measurement.} \textbf{Bottom:} Difference, in ${\rm cm}\ {\rm s}^{-1}$, of the standard deviation of the RV posterior distribution.
\label{Fig:mcmc_laplace_comp_timeseries}
\end{figure}

In order to compare RV estimates obtained through the MCMC methodology and the Laplace approximation we tested two targets, a K-type star with 27 ESPRESSO18 observations and a M-type star with 21 ESPRESSO19 observations, where ESPRESSO18 and ESPRESSO19 respectively refer to observations obtained before and after the ESPRESSO fiber link upgrade in June 2019 \citep{pepeESPRESSOVLTOnsky2021}. In Fig. \ref{Fig:mcmc_laplace_comp_timeseries} we show that the expected values for the RV shift obtained with the MCMC methodology and the Laplace approximation are, for both targets, within the respective uncertainties. The associated standard deviations are also in agreement at the ${\rm cm}\ {\rm s}^{-1}$ level, given that they differ at most by $\sim$ 2 ${\rm cm}\ {\rm s}^{-1}$ , with mean and median differences smaller than 0.3 ${\rm cm}\ {\rm s}^{-1}$), well below the expected 10 ${\rm cm}\ {\rm s}^{-1}$ precision of ESPRESSO.

\section{Results} \label{Sec:results}

In this Section we will showcase the results of the two template-based methodologies previously described by comparing them against the CCF method and two other template matching methods. For this purpose we use ESPRESSO data, reduced with version 2.2.8 of the official pipeline\footnote{\url{https://www.eso.org/sci/software/pipelines/}}, and HARPS archival data, reduced with version 3.5 of its official pipeline. We will use `classical' to refer to results obtained with the $\chi^2$ methodology discussed in Sect. \ref{Sec:chi2_method}, '\texttt{S-BART}' to refer to those from the semi-Bayesian methodology (coupled with the Laplace approximation) from Sect. \ref{Sec:Bayes_method}, and 'DRS' to those from the CCF of the instrument's pipeline. In order to assess the performance of our RV estimation methodologies we start by comparing our results with those from \texttt{HARPS-TERRA} \citep{anglada_escude_HARPS_TERRA_2012} and \texttt{SERVAL} \cite{zechmeisterSpectrumRadialVelocity2018} using the same 22 HARPS observations of Barnard's star. After that we focus our analysis on ESPRESSO data. In Sect. \ref{Sec:ESPRESSO_template_snr}, we select one ESPRESSO target and create multiple stellar templates, each from a different number of observations, to evaluate the impact on the RV scatter and median RV uncertainty. In Sect.  \ref{Sec:ESPRESSO_rms_comparison}, we select 33 ESPRESSO GTO targets and compare the scatter and precision of the two template-based methodologies (classical and \texttt{S-BART}) with those from the ESO pipeline (CCF). In Sect. \ref{Sec:RV_ESPRESSO_self_consistency} we evaluate whether our radial velocity uncertainties are consistent with the information present in the data, through the simulation of stellar spectra from one of the available observations. Lastly, in Sect. \ref{Sec:ESPRESSO_NZP} we use the same targets to estimate the nightly zero point (NZP) of the instrument with the three methodologies.

\subsection{Validation with HARPS data} \label{Sec:HARPS_results}

In order to validate our algorithm against other template matching methods we selected 22 HARPS observations of Barnard's star (GJ699), obtained between 2007-04-04 and 2008-05-02, with program ID 072.C-0488(E). This set of observations was chosen as it is present in the introductory papers of the two pipelines chosen for this purpose (\texttt{HARPS-TERRA} and \texttt{SERVAL}) and the observations are publicly available.

\begin{figure}[!ht]
\resizebox{\hsize}{!}{\includegraphics{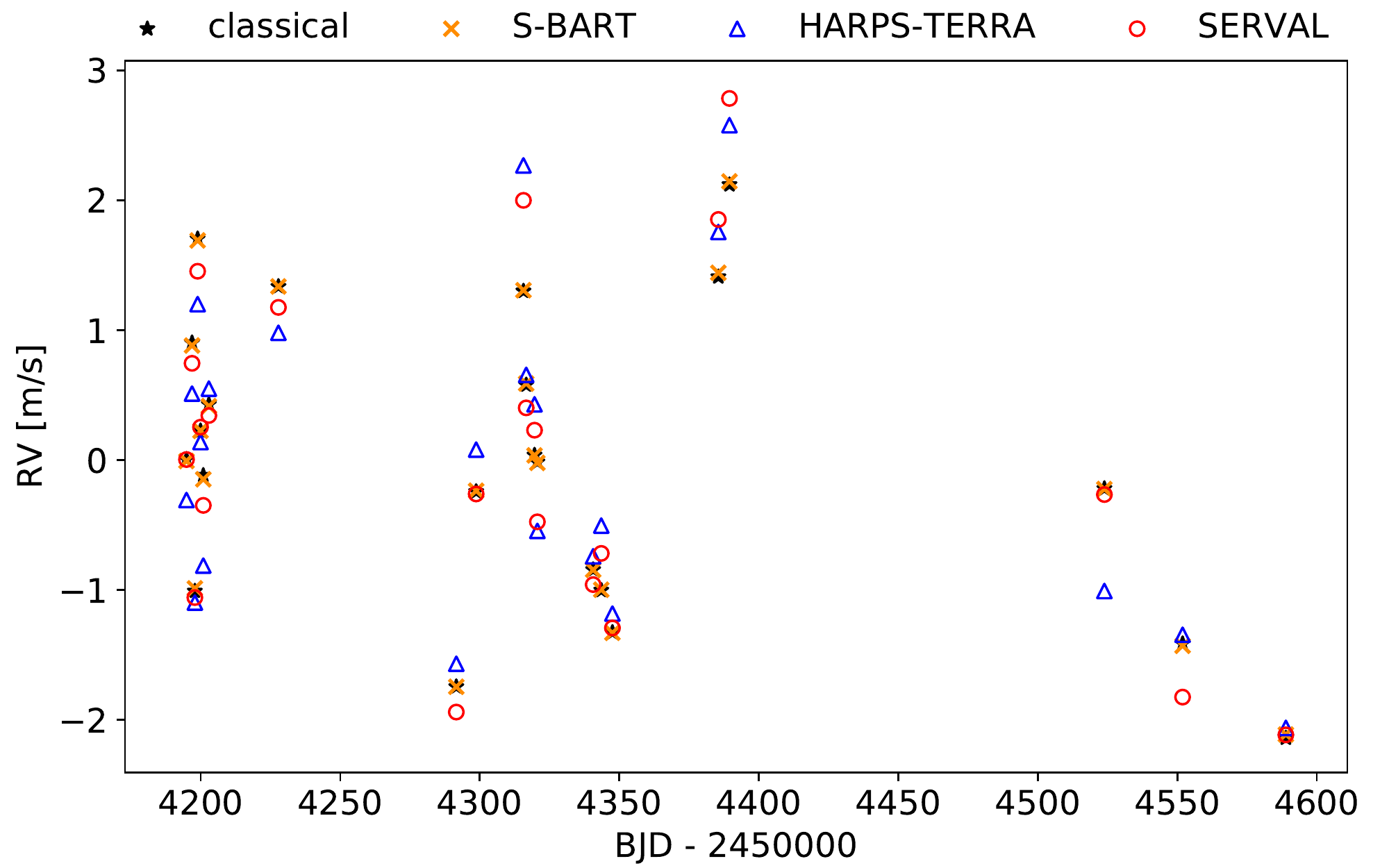}}
\caption{RV time-series for 22 observations of Barnard's star, corrected for the secular acceleration of the star. The black stars and orange crosses are RV estimates obtained with the classical and \texttt{S-BART} methodologies, respectively, with its own mean RV subtracted (for comparison purposes). The blue triangles and red dots are \texttt{HARPS-TERRA} and \texttt{SERVAL} RV estimates, respectively, also with their own mean RV subtracted.}
\label{Fig:HARPS_comparison}
\end{figure}

In Fig. \ref{Fig:HARPS_comparison} we  present the results obtained for \texttt{HARPS-TERRA}, \texttt{SERVAL} and our two template-matching methodologies. The \texttt{HARPS-TERRA} time-series was obtained from Table 6 of \citet{anglada_escude_HARPS_TERRA_2012}\footnote{We used the RVs obtained with the entire spectrum.} and the \texttt{SERVAL} time-series was derived by us using the most recent public version of \texttt{SERVAL}\footnote{\url{https://github.com/mzechmeister/serval}; commit d31a918}. For comparison purposes we show all RV estimates after subtracting the RV mean with respect to each method. A visual comparison of the different time-series allows to verify that the RV measurements show the same trends in all cases.

In Table \ref{Tab:HARPS_comparison} we compare the different methodologies with respect to the standard deviation of the RV estimates, a measure of scatter in the time-series. We include the RV scatter reported in \citep{zechmeisterSpectrumRadialVelocity2018}, as SERVAL-PAPER, for completeness. We find that both our methodologies reach the meter per second precision on HARPS data, achieving smaller scatter than with the \texttt{HARPS-TERRA} and \texttt{SERVAL} pipelines. The CCF-based \texttt{HARPS} pipeline leads to more scattered RV estimates, as expected given that Barnard's star belongs to the M spectral class. We refrain from comparing the estimates for the RV uncertainties as the different template-matching algorithms use different estimators for their calculation.

We find that our results are slightly less scattered than those obtained with other template-matching methods. Despite the small differences, they are concordant with the others, suggesting that our methodologies are working as intended. The same decrease in RV scatter is found for our two template-based RV time-series, suggesting that the different statistical framework is not the cause for such decrease. We believe it is due to differences in the way the stellar template is created and the telluric features are handled. In particular, \texttt{S-BART} attempts to minimize the impact of telluric features in RV estimation through a very conservative approach. This is achieved by creating a transmittance spectrum assuming the highest measured relative humidity amongst all observations of a given target, and then imposing a cut at 1\% transmittance. Lastly, it should be noted that the difference in RV scatter between \texttt{S-BART} and \texttt{HARPS-TERRA} is equal to that between the later and \texttt{SERVAL}.

\begin{table}[ht]
\caption{Time-series RV scatter obtained with different template-based methodologies when applied to Barnard's star.}
\label{Tab:HARPS_comparison}
\centering
\begin{tabular}{c c }
\hline\hline
Method &  std RV [${\rm m}\ {\rm s}^{-1}$]   \\
\hline
DRS-HARPS\tablefootmark{a}       &   1.51      \\
HARPS-TERRA \tablefootmark{b}  &   1.22      \\
SERVAL-PAPER \tablefootmark{b} &   1.30      \\
SERVAL \tablefootmark{a}       &   1.28      \\
classical \tablefootmark{c}       &   1.14      \\
\texttt{S-BART} \tablefootmark{c}         &   1.14      \\
\hline
\end{tabular}
\tablefoot{
\tablefoottext{a}{ The \texttt{SERVAL} and DRS-HARPS results were obtained by using the latest (publicly) available version of \texttt{SERVAL} and with the \texttt{HARPS} pipeline, respectively;}
 \tablefoottext{b}{The \texttt{HARPS-TERRA} and \texttt{SERVAL-PAPER} results were obtained from \citet{anglada_escude_HARPS_TERRA_2012} and \citet{zechmeisterSpectrumRadialVelocity2018}, respectively;}
 \tablefoottext{c}{Results obtained with the classical and \texttt{S-BART} methodologies.}
}
\end{table}

\subsection{Application to ESPRESSO data}

In this Section we compare of the performance of the CCF method of ESO's official pipeline with the application of the classical and \texttt{S-BART} methodologies, when applied to ESPRESSO data.

\subsubsection{Defining the stellar sample} \label{Sec:ESPRESSO_sample_def}

Our analysis of the performance of template matching uses data collected during 2018, 2019 and early 2020 (until March). The selected targets are part of ESPRESSO's blind RV survey program \citep{hojjatpanahCatalogESPRESSOBlind2019,pepeESPRESSOVLTOnsky2021} and all have at least 5 observations that can be used to construct a stellar template. ESPRESSO's fiber link was upgraded in June 2019 \citep{pepeESPRESSOVLTOnsky2021}, resulting in a change in the instrumental profile. We treat the data collected before and after the upgrade as if it was obtained from different instruments, i.e. we create independent stellar templates for the data obtained before and after the technical intervention. We shall refer to data obtained before and after the fiber link upgrade as `ESPRESSO18' and `ESPRESSO19', respectively.

In order to assess the performance of our template-based approaches we selected a sample of 33 targets, where 16 are M-type stars, 13 K-type stars and 4 G-type stars. In total, we used approximately 1000 observations distributed between ESPRESSO18 and ESPRESSO19, as specified in Table \ref{Tab:spectral_sample}.

\begin{table}[ht]
\caption{Number of observations, of each spectral type (ST), obtained before and after ESPRESSO's fiber link upgrade.}
\label{Tab:spectral_sample}
\centering
\begin{tabular}{c c c c c}
\hline\hline
ST & Targets & ESPRESSO18 & ESPRESSO19 & Total \\
\hline
M & 16 & 176 & 133 & 309 \\
K & 13 & 249 & 158 & 407 \\
G & 4 & 251 & 79 & 330 \\
\hline
\end{tabular}
\end{table}

The construction of the stellar template does not include any observation that has an airmass greater than 1.5, as discussed in Sect. \ref{Sec:build_template}. The selection of the targets in the sample was such that they all meet the condition, discussed in Sect. \ref{Sec:ESPRESSO_template_snr}, of having at least 5 observations that can be used in the construction of the stellar template.

\subsubsection{The impact of the number of observations in the template} \label{Sec:ESPRESSO_template_snr}

The first step to benchmark the performance of template-based RV estimation procedures with ESPRESSO data, and to understand for which targets we can take such an approach, is to evaluate the impact in the RV estimates of the number of observations used to construct the stellar template. For this purpose we selected, from the sample described in Sect. \ref{Sec:ESPRESSO_sample_def}, 24 ESPRESSO18 observations of an M-type star, from which we reserved the first 11 observations to construct stellar templates and used the other 13 to evaluate the performance of the templates. The 11 observations selected for the construction of the template cover a BERV region that starts at 25 ${\rm km}\ {\rm s}^{-1}$, in the first observation, and ends at -19 ${\rm km}\ {\rm s}^{-1}$ in the last one. The stellar templates are created by gradually selecting observations based on their BERV values, after they have been sorted from largest to smallest. Each template is then used to compute RVs for the aforementioned set of 13 observations. We do not use the same data to construct the template and to evaluate the performance of the RV estimation methods so that templates constructed with a low number of observations are not too similar to the spectra used to construct them.

\begin{figure}[!ht]
\resizebox{\hsize}{!}{\includegraphics{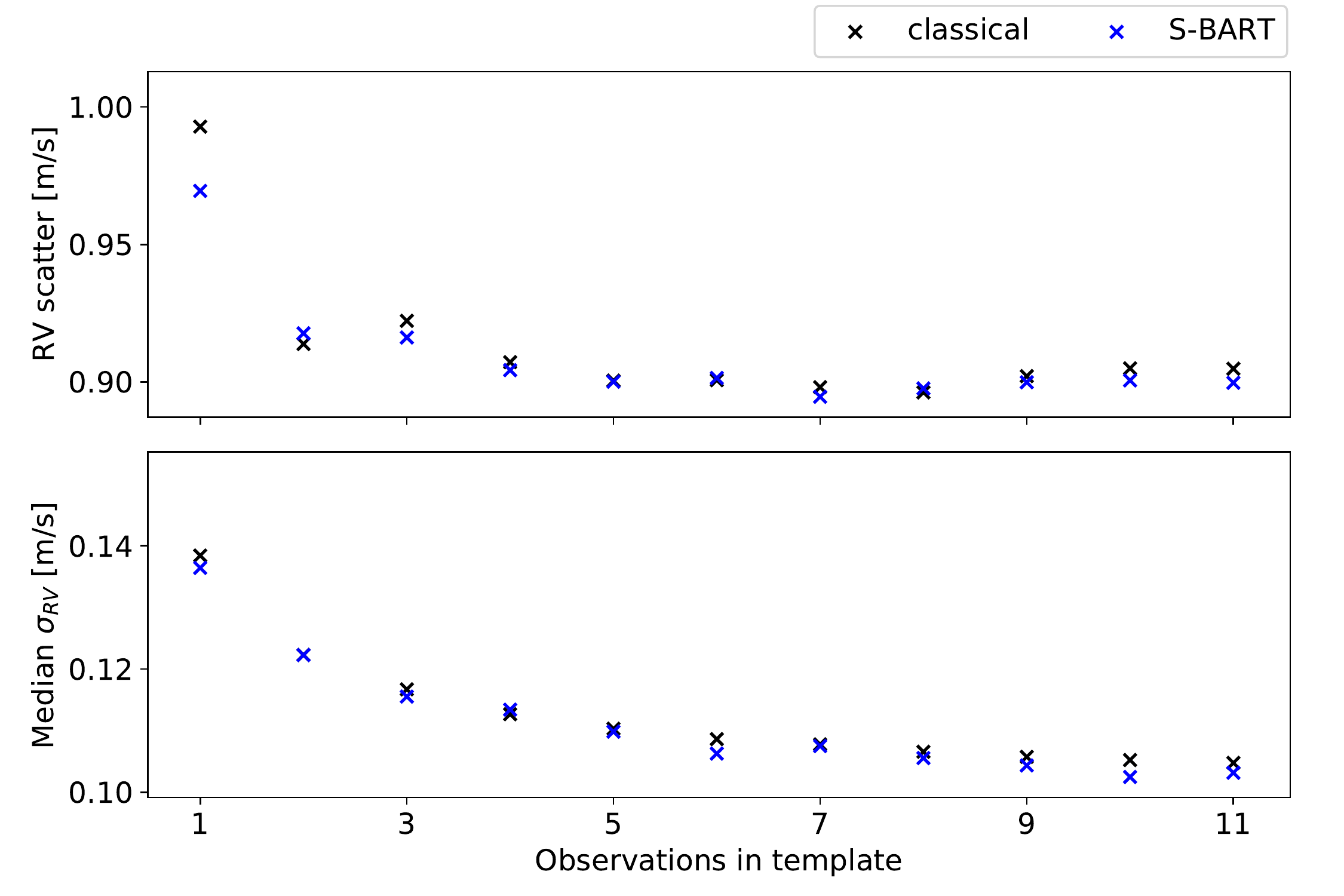}}
\caption{Evolution of the RV scatter (top pannel) and median uncertainty (bottom pannel) with respect to 13 ESPRESSO18 observations of a M-type star, as a function of the number of spectra used to construct the template. In black we have the results from the classical methodology and in blue those from the \texttt{S-BART} methodology. }
\label{Fig:template_evol_obs}
\end{figure}

We find an improvement in both the scatter and median uncertainty reported by both our methodologies as the SNR of the template increases (Fig. \ref{Fig:template_evol_obs}). If we focus only on the RV scatter, we find no meaningful improvement with templates constructed from more than 5 observations. We thus decided to estimate RVs only for targets that have at least 5 observations.

\subsubsection{Comparison of the RV scatter and precision} \label{Sec:ESPRESSO_rms_comparison}

We now compare the RV scatter and uncertainties obtained through the classical and \texttt{S-BART} methodologies. In order to achieve this, we  apply both of our template matching algorithms to the roughly 1000 observations that compose our ESPRESSO sample (Sect. \ref{Sec:ESPRESSO_sample_def}).

\begin{figure*}[ht]
\centering
\includegraphics[width=17cm]{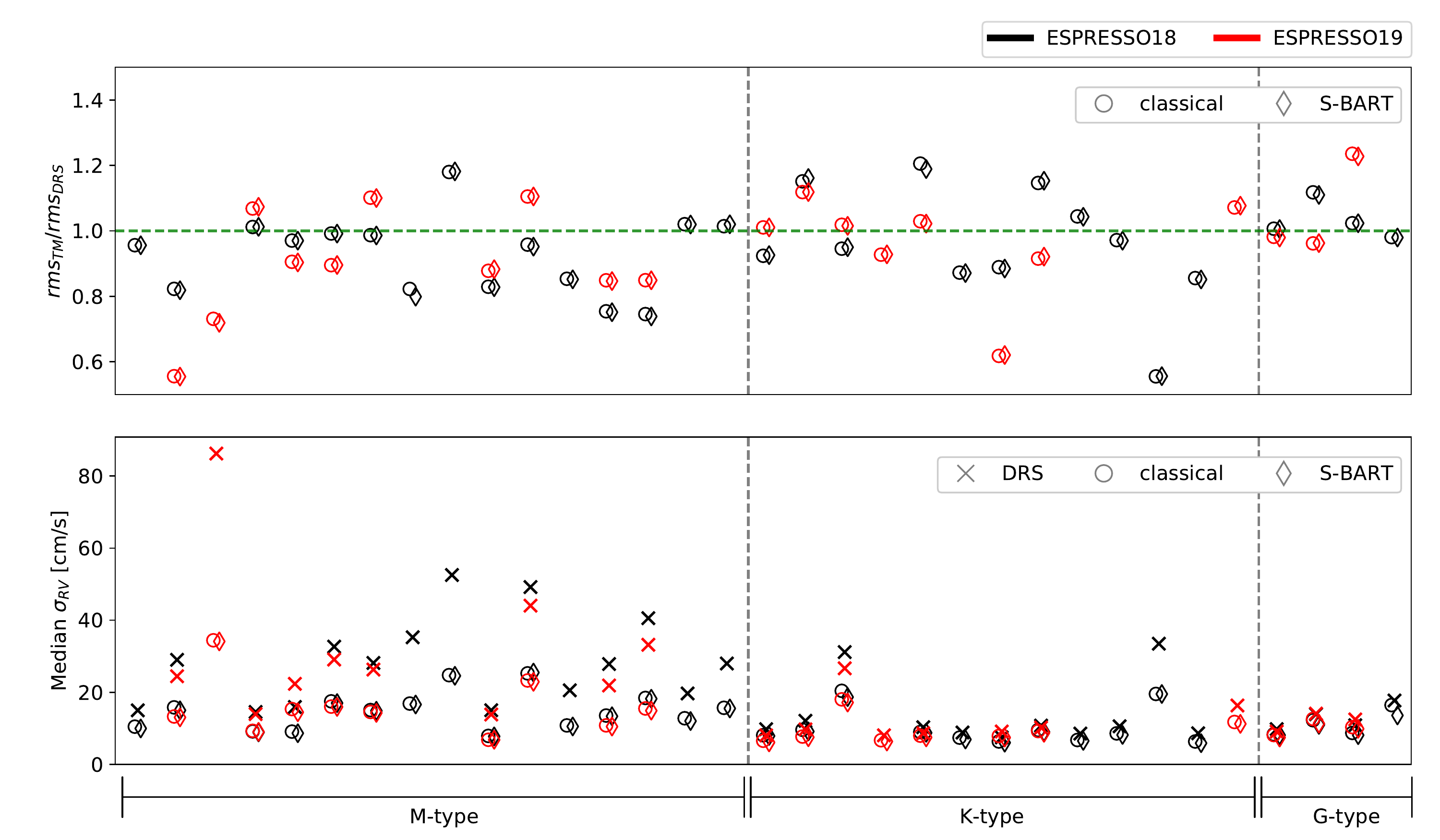}
\caption{Comparison of the results obtained with our template-matching methodologies and with the CCF-based ESPRESSO DRS, for ESPRESSO18 (black) and ESPRESSO19 (red) observations of the selected targets. \textbf{Top pannel:} Ratio of the rms in the template-matching and DRS RV time-series. The results from the classical and \texttt{S-BART} methods are represented by circles and diamonds, respectively. \textbf{Bottom pannel:} Median RV uncertainty for each target, as computed by the DRS (crosses), the classical approach (circle) and the \texttt{S-BART} method (diamonds).}
\label{Fig:target_comparison}
\end{figure*}

Similarly to other works, we find (Fig. \ref{Fig:target_comparison} and Table \ref{Tab:target_comparison_mean}) that the template-based methods, when applied to M-type stars, most often lead to a smaller scatter than the CCF method implemented in the DRS. This decrease is larger within the ESPRESSO19 dataset ($\sim$ 10\% smaller) than in the ESPRESSO18 one ($\sim$ 8\% smaller). For K-type stars we find a similar decrease across the two datasets, of $\sim$ 4\%. Lastly, for G-type stars the scatter, both before and after ESPRESSO fiber link upgrade, is $\sim$0.5\% larger than the DRS, a result that should be taken with caution due to the very limited sample size. The very similar RV estimates of the template-matching methodologies were expected, as both use the same information, i.e. the same spectral regions, and the same model (the template).

\begin{table}[ht]
\caption{Mean ratio between the scatter of the RV time series as derived through the two template-based methodologies and the CCF-based ESPRESSO DRS, separated by spectral type, ST, and methodology.}
\label{Tab:target_comparison_mean}
\centering
\begin{tabular}{ccccc}
\hline\hline
\noalign{\vskip 2pt}
Dataset                     & Method    & M-type & K-type & G-type \\
\hline
\noalign{\vskip 5pt}
\multirow{2}{*}{ESPRESSO18} & classical & 0.928  & 0.960  & 1.032  \\
                            & \texttt{S-BART}  & 0.923  & 0.961  & 1.029  \vspace{5pt} \\ \hline  \noalign{\vskip 5pt}
\multirow{2}{*}{ESPRESSO19} & classical & 0.894  & 0.964  & 1.060  \\
                            & \texttt{S-BART}  & 0.893  & 0.964  & 1.057 \\
\hline
\end{tabular}
\end{table}

\begin{figure*}[ht]
\centering
\includegraphics[width=17cm]{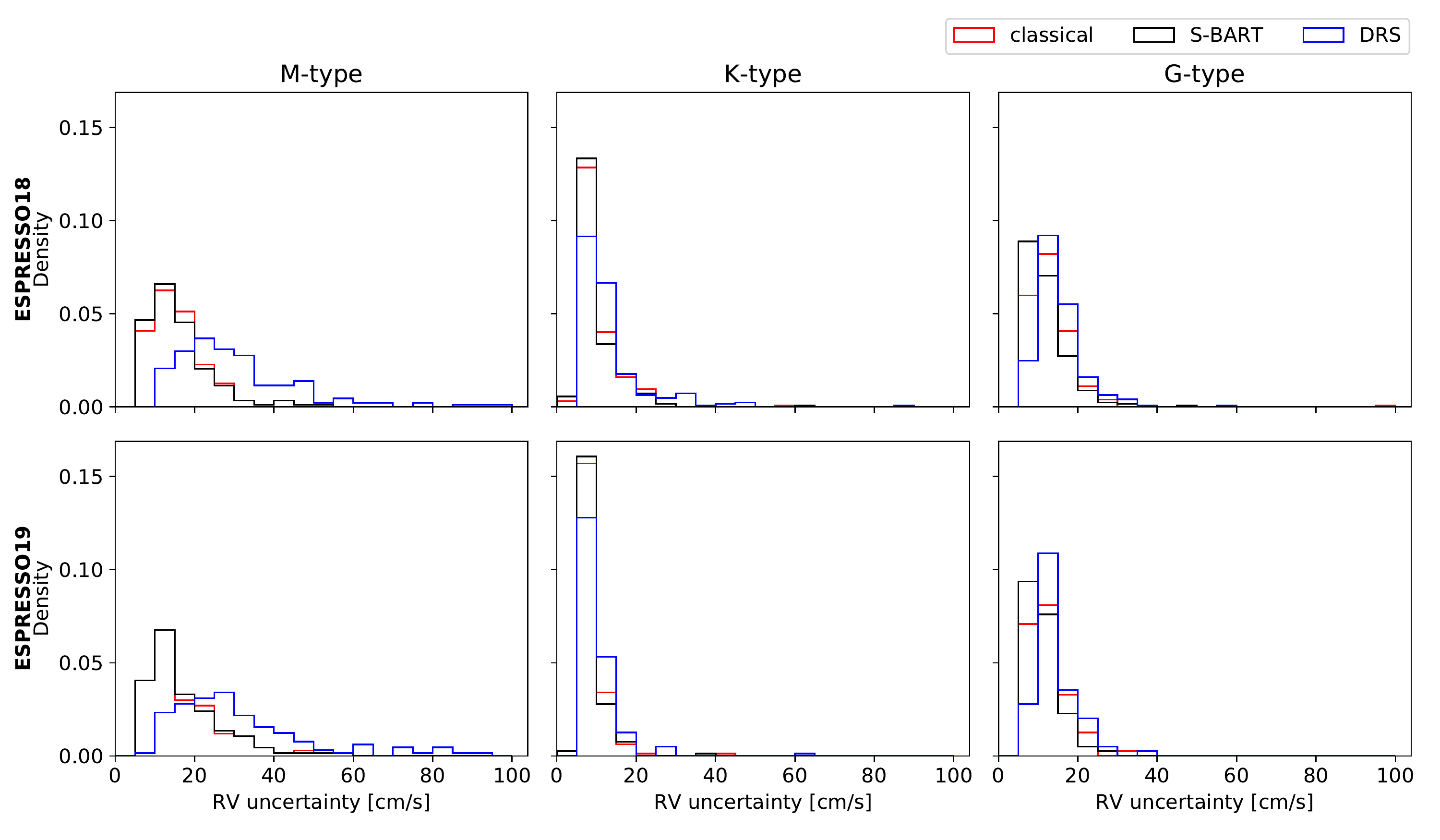}
\caption{Comparison of the RV uncertainties obtained for all ESPRESSO18 (top panel) and ESPRESSO19 (bottom panel) observations used in Figure \ref{Fig:target_comparison}, separated by spectral type. In blue we have the RV uncertainty estimated by the DRS, in black and red those estimated through the \texttt{S-BART} and classical methodologies, respectively.}
\label{Fig:espresso_error_comp}
\end{figure*}

We find very small differences, below the ${\rm cm}\ {\rm s}^{-1}$ mark, between the median RV uncertainties from the \texttt{S-BART} and classical approaches. Figure \ref{Fig:espresso_error_comp} shows the histograms of the individual RV uncertainty estimates for the observations, separated by spectral type, for each ESPRESSO dataset, whilst Table \ref{Tab:target_comparison_errors} summarizes the results. We see that for M-type stars both template-matching implementations yields a median RV uncertainty $\sim$ 13 ${\rm cm}\ {\rm s}^{-1}$ smaller than with the ESPRESSO DRS, corresponding to almost half of the median CCF RV uncertainty. For K- and G-type stars the gain in the median RV precision, in comparison with the CCF, is below 5 ${\rm cm}\ {\rm s}^{-1}$.

\begin{table}[ht]
\centering
\caption{Comparison of the median RV uncertainties, in ${\rm cm}\ {\rm s}^{-1}$, as obtained through the three methodologies.}
\label{Tab:target_comparison_errors}
\begin{tabular}{ccccc}
\hline\hline
\noalign{\vskip 2pt}
Dataset & Method & \multicolumn{1}{c}{M-type} & \multicolumn{1}{c}{K-type} & \multicolumn{1}{c}{G-type} \\
\hline
\noalign{\vskip 5pt}
\multirow{4}{*}{ESPRESSO18}&DRS & 26.7 & 10.4 & 13.4\\
&classical & 14.8 & 8.6 & 10.6\\
&\texttt{S-BART} & 14.3 & 8.1 & 10.6  \vspace{5pt} \\ \hline  \noalign{\vskip 5pt}
\multirow{4}{*}{ESPRESSO19}&DRS & 27.5 & 9.0 & 12.5\\
&classical & 14.7 & 7.6 & 10.9\\
&\texttt{S-BART} & 14.4 & 7.2 & 10.1\\
\hline
\end{tabular}
\tablefoot{ The values were derived for the two ESPRESSO datasets (ESPRESSO18 and ESPRESSO19) used in Fig. \ref{Fig:target_comparison}. }
\end{table}

We leave for future work an analysis of the impact in the RV estimates obtained using \texttt{S-BART} due to different levels of stellar activity. Such a complex endeavour lays outside the scope of the current paper, and will require a complete analysis of the RV time-series of each target, allowing for both stellar activity and (an unknown number of) planetary companions, It is also important to note that our current sample mostly contains stars with low levels of stellar activity \citep{hojjatpanahCatalogESPRESSOBlind2019}, meaning that we would either have to select a different stellar sample or be limited by that fact.

\subsubsection{Self-consistency of Radial Velocity uncertainties} \label{Sec:RV_ESPRESSO_self_consistency}

In order to determine whether our template matching estimates for the radial velocities uncertainties are consistent with the information present in the spectra, we simulated spectra and analysed it with the different methodologies:
\begin{enumerate}
    \item We start by selecting one reference observation;
    \item For each pixel in the reference spectrum, we draw a random value from a Gaussian distribution with mean and standard deviation equal to the flux value and uncertainty of the reference spectrum, respectively;
    \item We repeat the second step $N = 100$ times to build N 'simulated' spectra;
    \item We apply the template matching algorithm, using the stellar template that was built from the original data of the star whose observation was used as reference.
\end{enumerate}
We created two datasets from two observations of an M5 star, one from ESPRESSO18 and another from ESPRESSO19, and compared the uncertainty associated with the RV value estimated for the reference spectrum with the RV scatter and median RV uncertainty for the simulated dataset, as shown in Table \ref{Tab:white_noise_injectionTest}.
We find that for the three methodologies there is agreement between the RV uncertainty estimated for the original reference spectrum and the RV scatter and median uncertainty of the simulated datasets, even though for the assumed reference spectra the RV scatter and median uncertainties obtained with the CCF methodology are approximately double of those obtained with the \texttt{S-BART} and classical methods.

\begin{table}[h!]
\caption{Results, in ${\rm cm}\ {\rm s}^{-1}$, of the application of template matching and CCF to two simulated datasets where white noise was injected.}
\label{Tab:white_noise_injectionTest}
\begin{tabular}{ccccc}
\hline \hline
\multirow{2}{*}{ESPRESSO}                       & \multirow{2}{*}{method} & \multirow{2}{*}{$\sigma_{RV}$ reference}           & \multicolumn{2}{c}{Simulated data}        \\
                                                 &                         &   & std  & median $\sigma_{RV}$  \\ \hline \noalign{\vskip 3pt}
\multirow{3}{*}{18} & classical & 13.0   & 11.8          & 13.0                      \\
                    & \texttt{S-BART} & 12.8      & 11.8  & 12.9                      \\
                    & CCF      & 24.6                    & 26.2          & 24.6                 \vspace{1pt}     \\ \hline \noalign{\vskip 3pt}
\multirow{3}{*}{19} & classical       & 14.9          & 14.0     & 14.9                      \\
                    & \texttt{S-BART} & 14.8              & 14.1          & 14.9   \\
                    & CCF             & 28.0           & 29.3          & 28.0 \\  \hline
\end{tabular}
\tablefoot{One dataset was built from an ESPRESSO18 observation, whilst the other was from  ESPRESSO19 data, with both observations being from the same M5 star.}
\end{table}

On top of this analysis, we also made a comparison with the expected RV precision \citep{bouchyFundamentalPhotonNoise2001}, as implemented in \texttt{eniric} \citep{eniric_Neal2019}, revealing that the median \texttt{S-BART} uncertainties from each spectral type are just a few ${\rm cm}\ {\rm s}^{-1}$ above the corresponding photon noise limit.

\subsubsection{Nightly Zero Point (NZP) variation} \label{Sec:ESPRESSO_NZP}

The last study that we did with ESPRESSO data was an analysis of the nightly zero point (NZP) of each RV estimation procedure, following the methodologies implemented in \citet{courcolSOPHIESearchNorthern2015, tal-orCorrectingHIRESKeck2019}. For our analysis we again used the targets selected in Sect. \ref{Sec:ESPRESSO_sample_def} but, to enforce a balance of the number of observations between pre and post fiber link upgrade data, we do not consider G-type stars, as we only have 4 targets, with the majority of observations taken before the fiber link upgrade (Table \ref{Tab:spectral_sample}). Nonetheless, we still find that the ESPRESSO18 observations represent $\sim 60\%$ of our sample. It is important to note that our analysis uses a limited dataset that neither underwent a careful selection of targets nor had the contributions of stellar activity, planetary signals, and photon noise removed from the derived RVs. Thus, the subsequent results must be taken as an upper bound for the achievable stability of ESPRESSO.

The NZP calculation starts by subtracting, from the time-series of each target, its own error-weighted average, thus centering all time-series around an RV of 0 ${\rm m}\ {\rm s}^{-1}$.  If a target has multiple observations in the same night, we replaced them by the median value. We computed the NZP, for all nights in which at least 3 targets were observed, as the weighted average of the RVs, using weights equal to the inverse of the RV variances. The uncertainty in the NZP measurement is taken to be the maximum value between the propagated (through the weighted mean) RV uncertainty and the RV scatter of the night in question. For further details we refer back to the Appendix A of the original article.

\begin{figure*}[ht]
\centering
\includegraphics[width=17cm]{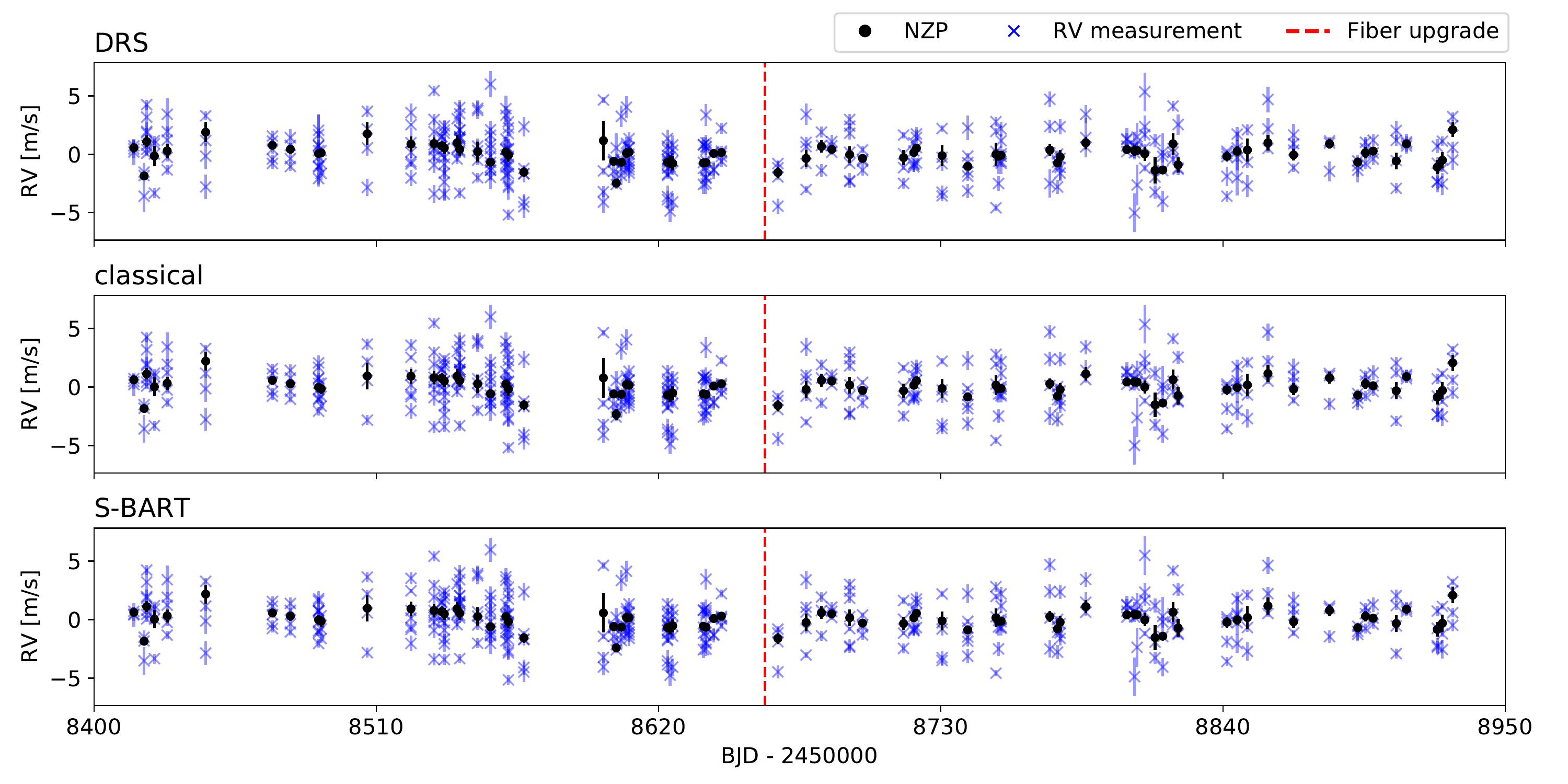}
\caption{Nightly Zero Point (NZP), black dots, for each night when at least 3 different targets were observed. The date of the fiber link upgrade is highlighted with a dashed red line. The zero-centered RV of each target, observed in the night, is represented with the blue crosses. As derived with results obtained with the following methodologies: ESPRESSO DRS (Top); classical (Middle); \texttt{S-BART} (Bottom).}
\label{Fig:NZP_comparison}
\end{figure*}

The NZP time-series is shown in Fig. \ref{Fig:NZP_comparison}, as derived from the RVs obtained with the ESPRESSO DRS as well as with our two template-matching methodologies. First, it is important to note that we have a higher density of targets per night in ESPRESSO18 than we do in ESPRESSO19. When visually comparing the NZPs obtained with the different RV estimation methods we find no significant differences, but an apparent smaller scatter in ESPRESSO19 data. This can be corroborated with a comparison of the weighted standard deviation of the NZPs (Table \ref{Tab:ESPRESSO_NZP}). We see that in both datasets the template-based results have a slightly lower scatter than those from the DRS, with the classical approach yielding the smallest NZP scatter, particularly with regards to ESPRESSO18 data. A comparison of ESPRESSO18 and ESPRESSO19 data reveals, across all methodologies, a weighted variability about 10 ${\rm cm}\ {\rm s}^{-1}$ smaller in the latter dataset.

\begin{table}[ht]
\caption{Weighted standard deviation, in ${\rm cm}\ {\rm s}^{-1}$, of the derived NZPs for observations obtained before and after the 2019 ESPRESSO fiber link upgrade.}
\label{Tab:ESPRESSO_NZP}
\centering
\begin{tabular}{ccc}
\hline\hline
\noalign{\vskip 2pt}
Method & ESPRESSO18 & ESPRESSO19 \\ \hline \noalign{\vskip 5pt}
ESPRESSO DRS &    76.0 &   59.9\\
classical &    69.6 &   57.2\\
\texttt{S-BART}   &   71.0 &   57.1\\
\hline
\end{tabular}
\tablefoot{ The results were obtained with the data from M and K-type stars, with RVs estimated through the three different methodologies.}
\end{table}

Despite the aforementioned limitations of our analysis, it is still noteworthy that, even under such circumstances, we find a NZP scatter below the meter per second mark, both before and after the fiber link upgrade.

\section{Limitations and possible improvements} \label{Sec:limitations}

The assumption that the stellar spectrum is invariant with time is a clear limitation in our template-matching approaches, which they share with other RV estimation procedures. We know that stellar activity induces spectral line displacements and deformations that change with time. This will thus induce a time-varying systematic error in the RV estimation, with a magnitude that is difficult to determine and that will depend on the star considered. This is the major obstacle to achieving precise, at the few tens of centimeter level, RV estimates. The solution will probably involve some form of data selection, namely at the spectral line level \citep{dumusqueMeasuringPreciseRadial2018,cretignierMeasuringPreciseRadial2020}, given that correctly modelling such complex effects seems a much more daunting task.

Even though the semi-Bayesian template-matching methodology, presented in Sect. \ref{Sec:Bayes_method}, improves the RV estimation with respect to the classical template-matching method, it has some shortcomings, that we are working on. First, the model for the spectra, effectively the prior with respect to the spectral fluxes, is built from the data that will be analysed, which means that we are computing the likelihood of the data with respect to a model that was built using information from the same data. The most straightforward solution to this problem would be to reserve a set of spectra solely for the purpose of constructing the template, but at the cost of losing RV estimates at the times those spectra were acquired. Another way of tackling this problem would involve using a probabilistic model for the assumed time-invariant true spectrum. This could involve physically relevant information, as in the CCF method, and has been implemented to some degree in \texttt{wobble} \citep{bedellWOBBLEDatadrivenAnalysis2019} and, using Gaussian Processes, in \texttt{GRACE} \citep{rajpaulRobustTemplatefreeApproach2020}.

Further, our semi-Bayesian approach depends on the classical implementation for the selection of the spectral orders to use as data. As discussed in the text, we are currently using those for which the classical template-matching procedure was able to obtain results, i.e. when our RV estimation convergence criteria were satisfied. However, it is possible that we could be discarding more (or less) information than we need to. One possible solution to this problem could be selection of orders through a maximization of the information gain between the (Gaussian) RV posterior and the (uniform) RV prior, using the Kullback-Leibler (KL) divergence \citep{kullbackInformationSufficiency1951}, which in this case is equivalent to minimizing the RV uncertainty (standard deviation). Similarly to what is done in the classical procedure, this approach would have to select the orders for all available observations simultaneously, but now such selection involves the full RV results, not only those calculated the level of the spectral order. Consequently, for targets with a large number of observations this procedure is laborious and compounds a major computational burden.

The approximation of the RV posterior probability distribution by a Gaussian using the Laplace approximation may not be the best procedure in some cases. This method only uses the information around the MAP estimate to build the approximation and, consequently, is unable to account for any skewness in the posterior. A more complex technique, such as Variational Inference \citep[for a detailed explanation refer to][]{gunapatiVariationalInferenceAlternative2022,bleiVariationalInferenceReview2017}, can use information from the entire posterior to build a more realistic approximation and, consequently, is more robust to skewness in the posterior.

\section{Conclusions}

In this work we revisited the template-matching approach for RV estimation in a semi-Bayesian framework, implemented in a pipeline names  S-BART:  Semi-Bayesian  Approach  for  RVs  with Template-matching. The key points of this approach are:
\begin{enumerate}
	\item The creation of an high-SNR stellar template with at least 5 observations with an airmass smaller than 1.5;
	\item A common RV shift is used to describe the differences between any given spectrum and a spectral template whose uncertainties are acounted for;
	\item The RV estimate and uncertainty are determined as the mean and standard deviation of the RV posterior probability distribution, respectively;
  	\item Due to the high computational cost of achieving convergence with an MCMC algorithm, we instead approximate the posterior with a Gaussian distribution, using Laplace's approximation.
\end{enumerate}

We compared the results of this new method with those obtained with the CCF approach and with a classical implementation of the template-matching algorithm, where independent RV shifts are assumed to describe the differences within each order between spectrum and template. The radial velocities are derived through the alignment of the spectra and a high signal-to-noise template, in which the uncertainties of the data used to construct it are considered.

In order to validate and evaluate the performance of our algorithm we applied it to observations from both HARPS and ESPRESSO, respectively. First, we compared the RV time-series obtained with our template-matching algorithms with those derived with the \texttt{HARPS-TERRA} and \texttt{SERVAL} pipelines, using 22 HARPS observations of Barnard's star. Our two methodologies yield a time series with a scatter $\sim$ 14 ${\rm cm}\ {\rm s}^{-1}$ smaller than the one from \texttt{SERVAL} and $\sim$ 8 ${\rm cm}\ {\rm s}^{-1}$ smaller than the one from \texttt{HARPS-TERRA}, revealing a good agreement between all RV estimates.

Afterwards we used \texttt{S-BART} to estimate RVs and the associated uncertainties for 33 ESPRESSO GTO targets of spectral types M, K and G. The median ratio between the RVs RMS obtained with our semi-Bayesian methodology and the CCF approach was 0.92 for M-type stars, 0.96 for K-type and 1.03 for G-type, for observations made before ESPRESSO's fiber link upgrade. After it, we obtain a median ratio of 0.89, 0.96 and 1.06 for M, K and G stars, respectively. The classical methodology also yielded similar results to those obtained with the \texttt{S-BART} method. This shows that the two template-matching approaches are able to provide more precise results for M-type stars, as one would expect, and also for K stars. Regarding the RVs uncertainties obtained with \texttt{S-BART}, we find a median value of $\sim$ 14 ${\rm cm}\ {\rm s}^{-1}$, $\sim$ 8 ${\rm cm}\ {\rm s}^{-1}$, $\sim$ 11 ${\rm cm}\ {\rm s}^{-1}$, for M-, K- and G-type stars, respectively. We left, for future work, a more detailed analysis of the signals, Keplerian or due to stellar activity, present in this sample.

Lastly, we also computed the nightly zero point (NZP) of the instrument, revealing a weighted NZP scatter around 0.7 ${\rm m}\ {\rm s}^{-1}$ for data obtained before the fiber link upgrade and 0.6 ${\rm m}\ {\rm s}^{-1}$ after it. Even though the scatter is higher than the expected precision of ESPRESSO, the NZPs were calculated without either removing stellar activity or planetary signals from the data and, consequently, should be taken as an upper limit of the obtainable precision.

\begin{acknowledgements}

The   authors   acknowledge   the ESPRESSO   project   team for   its   effort   and   dedication   in   building   the ESPRESSO  instrument. This  work  was  supported  by  FCT  - Fundação para   a   Ciência   e   a   Tecnologia   through national   funds   and   by   FEDER through  COMPETE2020  - Programa  Operacional  Competitividade  e  Inter-nacionalização by  these  grants: UID/FIS/04434/2019;   UIDB/04434/2020; UIDP/04434/2020;    PTDC/FIS-AST/32113/2017   \& POCI-01-0145-FEDER-032113; PTDC/FIS-AST/28953/2017    \& POCI-01-0145-FEDER-028953; PTDC/FIS-AST/28987/2017   \& POCI-01-0145-FEDER-028987.
A.M.S acknowledges support from the Fundação para a Ciência e a Tecnologia (FCT) through the Fellowship 2020.05387.BD. and POCH/FSE (EC).
J.P.F.  is  supported  in  the  form of  a  work  contract  funded  by  national  funds  through  FCT  with  reference DL57/2016/CP1364/CT0005.
S.G.S acknowledges the support from FCT through the contract nr. CEECIND/00826/2018 and POPH/FSE (EC)
J.H.C.M.  is  supported  in  the  form  of  a  work contract funded by national funds through FCT (DL 57/2016/CP1364/CT0007).
FPE and CLO would like to acknowledge the Swiss National Science Foundation (SNSF) for supporting research with ESPRESSO through the SNSF grants nr. 140649, 152721, 166227 and 184618. The ESPRESSO Instrument Project was partially funded through SNSF’s FLARE Programme for large infrastructures.
ASM, JIGH, and RR acknowledge financial support from the Spanish Ministry of Science and Innovation (MICINN)  project PID2020-117493GB-I00, and from the Government of the Canary Islands project ProID2020010129. JIGH also acknowledges financial support from the Spanish MICINN under 2013 Ram\'on y Cajal program RYC-2013-14875.
This project has received funding from the European Research Council (ERC) under the European Union’s Horizon 2020 research and innovation programme (project {\sc Four Aces}; grant agreement No 724427).  It has also been carried out in the frame of the National Centre for Competence in Research PlanetS supported by the Swiss National Science Foundation (SNSF). DE acknowledges financial support from the Swiss National Science Foundation for project 200021\_200726.
H.M.T. and M.R.Z.O ackowledge financial support from the Agencia Estatal de Investigaci\'on of the Ministerio de Ciencia, Innovaci\'on y Universidades through projects PID2019-109522GB-C51.
J.L-B. acknowledges financial support received from ”la Caixa” Foundation (ID 100010434) and from the European Union’s Horizon 2020 research and innovation programme under the Marie Skłodowska-Curie grant agreement No 847648, with fellowship code LCF/BQ/PI20/11760023.
R. A. is a Trottier Postdoctoral Fellow and acknowledges support from the Trottier Family Foundation. This work was supported in part through a grant from FRQNT. This work has been carried out within the framework of the National Centre of Competence in Research PlanetS supported by the Swiss National Science Foundation. The authors acknowledge the financial support of the SNSF.
The INAF authors acknowledge financial support of the Italian Ministry of Education, University, and Research with PRIN 201278X4FL and the "Progetti Premiali" funding scheme.
V.A. acknowledges the support from FCT through Investigador FCT contract nr. IF/00650/2015/CP1273/CT0001;
NJN acknowledges support form the following projects:
CERN/FIS-PAR/0037/2019, PTDC/FIS-OUT/29048/2017;
Based on data products from observations made with ESO Telescopes at the La Silla Paranal Observatory
under programme ID 072.C-0488(E).
\end{acknowledgements}

\bibliographystyle{aa} 
\bibliography{biblio}

\begin{appendix}
\section{Uncertainty propagation in cubic splines} \label{App:error_propagation}
The cubic-spline interpolation, used to calculate a value \textit{y} at position \textit{x} in the interval [$x_i, x_{i+1}$], can be written in the following form \citep[p. 113 of ][]{pressNumericalRecipesArt1992} :

\begin{equation} \label{Eq:spline_equation}
y = Ay_i + By_{i+1} + C y^{''}_{i} + D y^{''}_{i+1}
\end{equation}

\noindent where the double apostrophe represents the second derivative with respect to \textit{x} and

\begin{equation}
\begin{split}
A = \frac{x_{i+1} - x}{x_{i+1} - x_i}\\
B = \frac{x - x_i}{x_{i+1} - x_i}\\
C = \frac{1}{6}(A^3 - A)(x_{i+1} - x_i)^2\\
D = \frac{1}{6}(B^3 - B)(x_{i+1} - x_i)^2\\
\end{split}
\end{equation}

The computation of the second derivatives requires choosing proper boundary conditions. As discussed in Sect. \ref{Sec:build_template} we remove the wavelength regions that are not common to all observations. Thus, as we do not interpolate in the edges, we can use natural boundary conditions, where the second derivative takes a value of zero at the edges of the input data. Following the notation of \citet{pressNumericalRecipesArt1992} we can write down a general expression for the second derivatives:

\begin{equation} \label{Eq:derivative_expr}
\begin{split}
\begin{pmatrix}
\frac{y_{3} - y_2}{x_{3} - x_2} - \frac{y_2-y_{1}}{x_2-x_{1}} \\
\vdots \\
\frac{y_{i+1} - y_i}{x_{i+1} - x_i} - \frac{y_i-y_{i-1}}{x_i-x_{i-1}}\\
\vdots \\
\frac{y_{N} - y_{N-1}}{x_{N} - x_{N-1}} - \frac{y_{N-1}-y_{N-2}}{x_{N-1}-x_{N-2}} \\
\end{pmatrix} = h_{i,j}\begin{pmatrix}
y^{''}_2 \\
\vdots \\
y^{''}_i \\
\vdots  \\
y^{''}_{N-1} \\
\end{pmatrix} \\
y^{''}_1 = y^{''}_N = 0
\end{split}
\end{equation}

\noindent where h is a symmetric tridiagonal matrix given by:

\begin{equation}
h_{i,j} = \begin{pmatrix}
\frac{x_{3} - x_{1}}{3} & \frac{x_{3} - x_{2}}{6} & 0 & 0 \\
\frac{x_i - x_{i-1}}{6} & \frac{x_{i+1} - x_{i-1}}{3} & \frac{x_{i+1} - x_{i}}{6} & 0 \\
0 & \ddots & \ddots & \ddots & \\
\\
0 & 0 & \frac{x_{N-1} - x_{N-2}}{6} & \frac{x_{N} - x_{i-2}}{3} \\
\end{pmatrix}
\end{equation}

An expression for the propagation of uncertainties in cubic splines was derived in \citet{gardnerUncertaintiesInterpolatedSpectral2003}, with the covariance between any two given interpolated points, $u(y_m,y_n)$, given by:

\begin{equation} \label{Eq:simplified_U}
\begin{split}
u(y_m,y_n) = \begin{pmatrix} A_i &B_i&C_i&D_i\end{pmatrix} \\
\begin{pmatrix}
u(y_i,y_j) & u(y_i,y_{j+1}) & u(y_i,y^{''}_j) & u(y_i,y^{''}_{j+1}) \\
u(y_{i+1},y_j) & u(y_{i+1},y_{j+1}) & u(y_{i+1},y^{''}_j) & u(y_{i+1},y^{''}_{j+1}) \\
u(y^{''}_i,y_{j})&u(y^{''}_i,y_{j+1}) & u(y^{''}_i,y^{''}_j) & u(y^{''}_i,y^{''}_{j+1}) \\
u(y^{''}_{i+1},y_{j})&u(y^{''}_{i+1},y_{j+1}) & u(y^{''}_{i+1},y^{''}_j) & u(y^{''}_{i+1},y^{''}_{j+1}) \\
\end{pmatrix}
\begin{pmatrix} A_j \\B_j\\C_j\\D_j\end{pmatrix}
\end{split}
\end{equation}

\noindent with

\begin{equation}
\begin{split}
u(y^{''}_m, y^{''}_n) = Q^T_m U_y Q_n \\
u(y^{''}_m, y_n) = Q^T_m U_y g_n
\end{split}
\end{equation}

\noindent and where $U_y$ is a N$\times$N covariance matrix of the input data, $g_n$ is a column vector of length N with a 1 in the n-th column and 0 elsewhere, $Q_i$ is a column-vector of sensitivity coefficients of the second-derivatives in relation to the input values

\begin{equation} \label{Eq:f_description}
Q_m =\begin{cases}
\frac{\partial y_m^{''}}{\partial y_k} & k \in \{2,...,N-1\} \\
0 & {\rm otherwise}
\end{cases}
\end{equation}

In order to calculate the partial derivative (of the second derivatives) we start by selecting a given row from Equation \ref{Eq:derivative_expr} and re-arranging the summation limits:

\begin{equation} \label{Eq:second_der_final_form}
\begin{split}
y^{''}_i = -\sum^{N-1}_{j=2} h_{i,j}^{-1} \left( \frac{1}{x_{j+1} - x_j} + \frac{1}{x_j - x_{j-1}}\right) y_j +\\+ \sum^{N}_{j=3} h_{i,j-1}^{-1} \frac{y_{j}}{x_{j} - x_{j-1}} + \sum^{N-2}_{j=1} h_{i,j+1}^{-1}
\frac{y_{j}}{ x_{j+1} - x_{j} }
\end{split}
\end{equation}

The partial derivative then follows from Equation \ref{Eq:second_der_final_form}:

\begin{equation} \label{Eq:second_derivative_general}
\begin{split}
\frac{\partial y_i^{''}}{\partial y_j} = - h_{i,j}^{-1} \left( \frac{1}{x_{j+1} - x_j} + \frac{1}{x_j - x_{j-1}}\right) +\\+  \frac{h_{i,j-1}^{-1}}{x_{j} - x_{j-1}} + \frac{ h_{i,j+1}^{-1}}{ x_{j+1} - x_{j} }
\end{split}
\end{equation}

Special care must be taken due to the summation limits in Equation \ref{Eq:second_der_final_form}, as not all terms in Equation \ref{Eq:second_derivative_general} exist for all indexes. As we are only interested in the variances of the interpolated values, i.e. we will not consider the effect of the covariances, we only have to evaluate Equation \ref{Eq:simplified_U} for the $m = n$ case.

Lastly, as we are mainly dealing with ESPRESSO data, the computation of the second derivatives through Eq. \ref{Eq:derivative_expr} implies the inversion of a matrix with size $N-2\times N-2$, with $N = 9111$, which poses a large computational burden. However, as we are dealing with a symmetric tridiagonal matrix of the form:

\begin{equation}
h = \begin{pmatrix} x_1 & y_1 \\
z_1 & x_2 & \ddots \\
& \ddots & \ddots & y_{N-1} \\
& & z_{N-1} & y_N
\end{pmatrix}
\end{equation}

\noindent where $z_n$ = $y_n$, we can invert it with an explicit formula, using backwards continued fractions \citep{kilicExplicitFormulaInverse2008}:

\begin{equation}
h^{-1}_{ij} =
\begin{cases}
\displaystyle\frac{1}{C^b_i} + \sum^{N}_{k=i+1}\left(\frac{1}{C^b_k} \prod^{k-1}_{t=i} \frac{y^2_t}{(C^b_t)^2} \right)  & {\rm if}\ i = j \\
\\
\displaystyle (-1)^{i+j} \prod^{i-1}_{t=j} \frac{y_t}{C^b_t}h^{-1}_{ii} & {\rm otherwise}
\end{cases}
\end{equation}

\noindent with $C^b_n$ given by Eq 4 of the aforementioned paper:

\begin{equation}
C_n^b =
\begin{cases}
x_1 & {\rm if}\ n = 1 \\
x-n + \frac{-y_{n-1}z_{n-1}}{C^b_{n-1}} & {\rm otherwise}
\end{cases}
\end{equation}

\end{appendix}

\end{document}